\documentclass[aps,10pt,pra,superscriptaddress,showpacs,floatfix,twocolumn,longbibliography,nofootinbib]{revtex4-1}

\usepackage{graphicx}
\usepackage{epstopdf}
\usepackage{amsmath}
\usepackage{comment}
\usepackage{amssymb}
\usepackage{hyperref}
\usepackage{bm}
\usepackage{subfigure}

\hypersetup{
    colorlinks=true,       
    linkcolor=cyan,          
    citecolor=magenta,        
    filecolor=magenta,      
    urlcolor=cyan,           
    runcolor=cyan
}
\usepackage[pdftex]{color}

\usepackage{threeparttable}
\usepackage{subfigure}
\usepackage{upgreek }
\usepackage{marginnote}
\usepackage{cancel}

\newcommand{\beq}{\begin{equation}}
\newcommand{\eeq}{\end{equation}}
\newcommand{\beqnn}{\begin{equation*}}
\newcommand{\eeqnn}{\end{equation*}}
\newcommand{\bea}{\begin{eqnarray}}
\newcommand{\eea}{\end{eqnarray}}
\newcommand{\beann}{\begin{eqnarray*}}
\newcommand{\eeann}{\end{eqnarray*}}
\newcommand{\bes} {\begin{subequations}}
\newcommand{\ees} {\end{subequations}}

\newcommand{\ket}[1]{ | #1\rangle}
\newcommand{\bra}[1]{\langle #1 | }
\newcommand{\ketbra}[2]{|#1\rangle\langle #2|}

\newcommand{\ident}{\openone}

\newcommand{\ignore}[1]{}

\begin{document}
\title{Diagonal Catalysts in Quantum Adiabatic Optimization} 

\author{Tameem Albash}
\affiliation{Department of Electrical and Computer Engineering,  University of New Mexico, Albuquerque, New Mexico 87131, USA}
\affiliation{Department of Physics and Astronomy and Center for Quantum Information and Control, CQuIC, University of New Mexico, Albuquerque, New Mexico 87131, USA}

\author{Matthew Kowalsky}
\affiliation{Department of Physics and Astronomy, University of Southern California, Los Angeles, California 90089, USA}
\affiliation{Center for Quantum Information Science \& Technology, University of Southern California, Los Angeles, California 90089, USA}

\begin{abstract}
We propose a protocol for quantum adiabatic optimization, whereby an intermediary Hamiltonian that is diagonal in the computational basis is turned on and off during the interpolation.  This `diagonal catalyst' serves to bias the energy landscape towards a given spin configuration, and we show how this can remove the first-order phase transition present in the standard protocol for the ferromagnetic $p$-spin and the Weak-Strong Cluster problems. The success of the protocol also makes clear how it can fail: biasing the energy landscape towards a state only helps in finding the ground state if the Hamming distance from the ground state and the energy of the biased state are correlated.  We present examples where biasing towards low energy states that are nonetheless very far in Hamming distance from the ground state can severely worsen the efficiency of the algorithm compared to the standard protocol.  Our results for the diagonal catalyst protocol are analogous to results exhibited by adiabatic reverse annealing, so our conclusions should apply to that protocol as well.  
\end{abstract}

\maketitle
\section{Introduction}

The ubiquity of optimization problems continues to nurture the study and development of new algorithms to reduce the computational cost of solving them. Quantum adiabatic optimization \cite{finnila_quantum_1994,Brooke1999,kadowaki_quantum_1998,Farhi:01,Santoro} (QAO) is an approach that uses a sufficiently slow quantum evolution subject to an interpolating Hamiltonian $H(s), s \in [0,1],$ to find the solution to the optimization problem encoded in the ground state of the problem Hamiltonian $H_\mathrm{P}$ at the end of the interpolation, $H(1) = H_{\mathrm{P}}$. Recent progress in the development of large-scale qubit systems \cite{harris_flux_qubit_2010,Johnson:2010ys,Berkley:2010zr,Bunyk:2014hb} has increased the impetus to better understand when we can expect QAO to provide a quantum advantage over classical algorithms for real-world optimization problems. To date, such advantages are only known in the oracular setting \cite{Roland:2002ul,Has2020}\footnote{The exponential speedup in Ref.~\cite{Somma:2012kx} is not strictly speaking in the adiabatic setting.}.

In an analogous manner to its classical counterpart \cite{kirkpatrick_optimization_1983}, QAO relies on the adiabatic theorem of quantum mechanics \cite{Einstein:adiabatic,Ehrenfest:adiabatic,Born:28,Kato:50,Jansen:07} to provide a guarantee that evolutions satisfying the adiabatic condition will have a high overlap with the ground state of $H_{\mathrm{P}}$.  The condition on how slowly the evolution must be performed is usually stated as a condition on the total evolution time $t_f$, whereby $t_f$ must be much greater than some power of the inverse minimum energy gap $\Delta_\mathrm{min}^{-1}$ of $H(s)$ along the interpolation.  This then gives a convenient way to express the computational time-cost of the algorithm, with the scaling of the minimum gap with system size $n$ often reported as representing the efficiency of the QAO algorithm.

In the standard (`S') QAO setting, the interpolating Hamiltonian is written as a linear combination of a  `driver' Hamiltonian $H_{\mathrm{D}}$ and a `problem' Hamiltonian  $H_{\mathrm{P}}$ that is diagonal in the computational basis:
\beq \label{eqt:H_S}
H_\mathrm{S} (s) = (1-s) H_{\mathrm{D}} + s H_{\mathrm{P}} \ ,
\eeq
where for simplicity we have taken a linear interpolation for the annealing schedule.
We take the single qubit computational basis $\left\{\ket{0}, \ket{1}\right\}$ to be the eigenstates of the Pauli-$z$ operator, $\sigma^z \ket{0} = \ket{0}$, $\sigma^z \ket{1} = - \ket{1}$, corresponding to the spin-up and spin-down states respectively.  The driver Hamiltonian is usually taken to be the uniform transverse field Hamiltonian $H_{\mathrm{D}} = - \sum_i \sigma_i^x$.

Increased experimental control capabilities of quantum annealing systems \cite{King2018,Harris162} has lead to a resurgence of interest in different interpolation paths than those used in standard quantum annealing.  Notably, there has been renewed interest in `adiabatic reverse annealing' \cite{PhysRevA.98.022314,2019arXiv190610889Y} (ARA), or `sombrero Adiabatic Quantum Computing (AQC)' \cite{Perdomo-Ortiz:2011fh}, whereby the interpolation starts from a diagonal local-field Hamiltonian $H_{\mathrm{B}} = -\sum_i \varepsilon_i \sigma_i^z$ that encodes a classical spin configuration $\vec{\varepsilon} \in \left\{-1,1 \right\}^n$ as its ground state.  The spin state $\vec{\varepsilon}$ will generically disagree with the ground state of $H_{\mathrm{P}}$ over some subset of the indices $i$. We denote the fraction of spins that agree with the ground state of $H_{\mathrm{P}}$ by $c$.
 
Studying the performance of ARA on the $p$-spin model, Ref.~\cite{PhysRevA.98.022314} found that above a critical value of $c$, the scaling of the minimum gap changes from exponentially closing to only polynomially closing, indicating an exponential improvement in the performance of the algorithm.  These results have been taken as a positive indication that greater control of and choices for the interpolating Hamiltonian may result in dramatically improved performance for solving hard optimization problems.

Here we propose an alternative interpolating path using the same terms of the ARA Hamiltonian, $\left\{H_{\mathrm{D}}, H_{\mathrm{B}}, H_\mathrm{P} \right\}$, but that is more akin to the standard interpolation in that the initial Hamiltonian is the driver Hamiltonian:
\beq \label{eqt:H_DC}
H_{\mathrm{DC}}(s) = (1-s) H_{\mathrm{D}} + \lambda s (1-s)  H_{\mathrm{B}} + s H_{\mathrm{P}} \ .
\eeq
In the new interpolation, the Hamiltonian $H_{\mathrm{B}}$ is introduced as a `catalyst' Hamiltonian \cite{RevModPhys.90.015002,Seki:2012,2018arXiv180607602D,nonStoq} that is turned on and off during the interpolation.  Because this catalyst is diagonal in the computational basis, we refer to it as a `diagonal catalyst' (DC). This new approach reproduces the exponential improvement in the performance of solving the $p$-spin model over the standard QAO algorithm, with the added benefit that it makes clear the role of the catalyst $H_{\mathrm{B}}$ in providing the performance improvement: it works by biasing the energy landscape towards the target solution.

Our analysis also highlights two important limitations of this approach.  First, as the value of $p$ increases, the overall strength of the catalyst Hamiltonian must be made larger, and in the limit of $p \to \infty$, where the $p$-spin model becomes similar to the problem of unstructured search \cite{Roland:2002ul,Jorg:2010qa}, maintaining an exponential improvement requires the overall strength to increase linearly with system size.  

Second, it becomes clear that biasing the energy landscape only works if we are biasing \emph{towards} the target state, where the relevant distance measure is Hamming distance and not how close they are in energy.  We highlight this by constructing instances where the low-lying energy states of $H_{\mathrm{P}}$ are far in Hamming distance from the ground state, and biasing towards these states makes the performance of the algorithm significantly worst than the standard protocol.

Our paper is structured as follows. In Sec.~\ref{sec:pSpin}, we study the ferromagnetic $p$-spin model, and show how a suitable choice of the diagonal catalyst can eliminate the first-order phase transition associated with this model.  In Sec.~\ref{sec:LargeP}, we study the large $p$ limit, corresponding to the problem of unstructured search and show how the diagonal catalyst fails to eliminate the first-order phase transition in this model unless the catalyst becomes infinitely strong.  In Sec.~\ref{sec:WeakStrong}, we study the Weak-Strong Cluster problem \cite{GoogleTunneling}, which is another example where the first-order phase transition can be eliminated by a suitably chosen diagonal catalyst.  In Sec.~\ref{sec:Bottleneck}, we present a failure mechanism of the protocol and show how the diagonal catalyst can exacerbate or introduce new bottlenecks to the standard protocol.  In Sec.~\ref{sec:Conclusions}, we provide a discussion and concluding remarks about our protocol and its relationship to adiabatic reverse annealing.
\section{Fully-connected ferromagnetic $p$-spin model} \label{sec:pSpin}
We begin our analysis of the performance of our DC interpolation (Eq.~\eqref{eqt:H_DC}) by studying the fully-connected ferromagnetic $p$-spin model
\beq \label{eqt:pSpin}
H_{\mathrm{P}} = -n\left(\frac{1}{n} \sum_{i} \sigma_{i}^{z}\right)^{p} \ .
\eeq
The ground state of this problem Hamiltonian is the all-zero state, $\ket{0}^{\otimes n}$, corresponding to a magnetization density of 1.  In the standard interpolation (Eq.~\eqref{eqt:H_S}), the minimum gap along the interpolation closes exponentially with system size \cite{Jorg:2010qa}, and in the thermodynamic limit the closing of the gap is associated with a first-order phase transition.    

\subsection{Mean-field analysis}
We first perform a mean-field analysis of our DC Hamiltonian in Eq.~\eqref{eqt:H_DC}.  The mean field analysis and the resulting free energy density in terms of the mean-field magnetization density provides a simple characterization of the energy landscape during the interpolation.  Discontinuous jumps in the identity of the global minimum of the free energy are associated with a first-order phase transition, since the magnetization changes discontinuously when following the global minimum.  Following the derivation of Ref.~\cite{PhysRevA.98.022314}, the mean-field free energy in the zero-temperature limit can be written as
\begin{eqnarray} \label{eqt:pSpinFreeEnergy}
f(m) &=& s (p-1) m^p \nonumber \\
&& \hspace{-1cm}- c \sqrt{ s^2 \left( p m^{p-1} + \lambda (1-s)\right)^2 + (1-s)^2 } \nonumber \\
&& \hspace{-1cm} - (1-c)  \sqrt{ s^2 \left( p m^{p-1} - \lambda (1-s) \right)^2 + (1-s)^2 } \ , \label{eqt:f} 
\end{eqnarray}
where $m$ denotes the mean-field magnetization density.  For $\lambda = 0$, corresponding to the absence of the DC, the free energy density reverts to that of the standard interpolation. We identify the location of the first-order phase transition along the interpolation by solving for the global minimum of the free energy $f(m)$ and identifying the value of the interpolation parameter $s = s_\ast$ where a discontinuous jump in the global minimum occurs.  

 In the absence of a catalyst, $f(m)$ exhibits a degenerate double-well at some point along the interpolation.  When the catalyst is turned on, the separation of the degenerate minima is reduced.  Above a critical value $c$ and when the catalyst is sufficiently strong, the two minima merge and the double well vanishes, and only a single minimum is realized during the entire anneal. This behavior is qualitatively similar to that observed for the ARA protocol \cite{PhysRevA.98.022314}. Here, the role of the catalyst is to energetically bias the landscape towards $m=1$, pushing the first minimum towards the second minimum until only one minimum remains.  Further details are provided in Appendix~\ref{app:Landscape}. 
 
\begin{figure}[tbp] 
   \centering
\includegraphics[width=0.75\columnwidth]{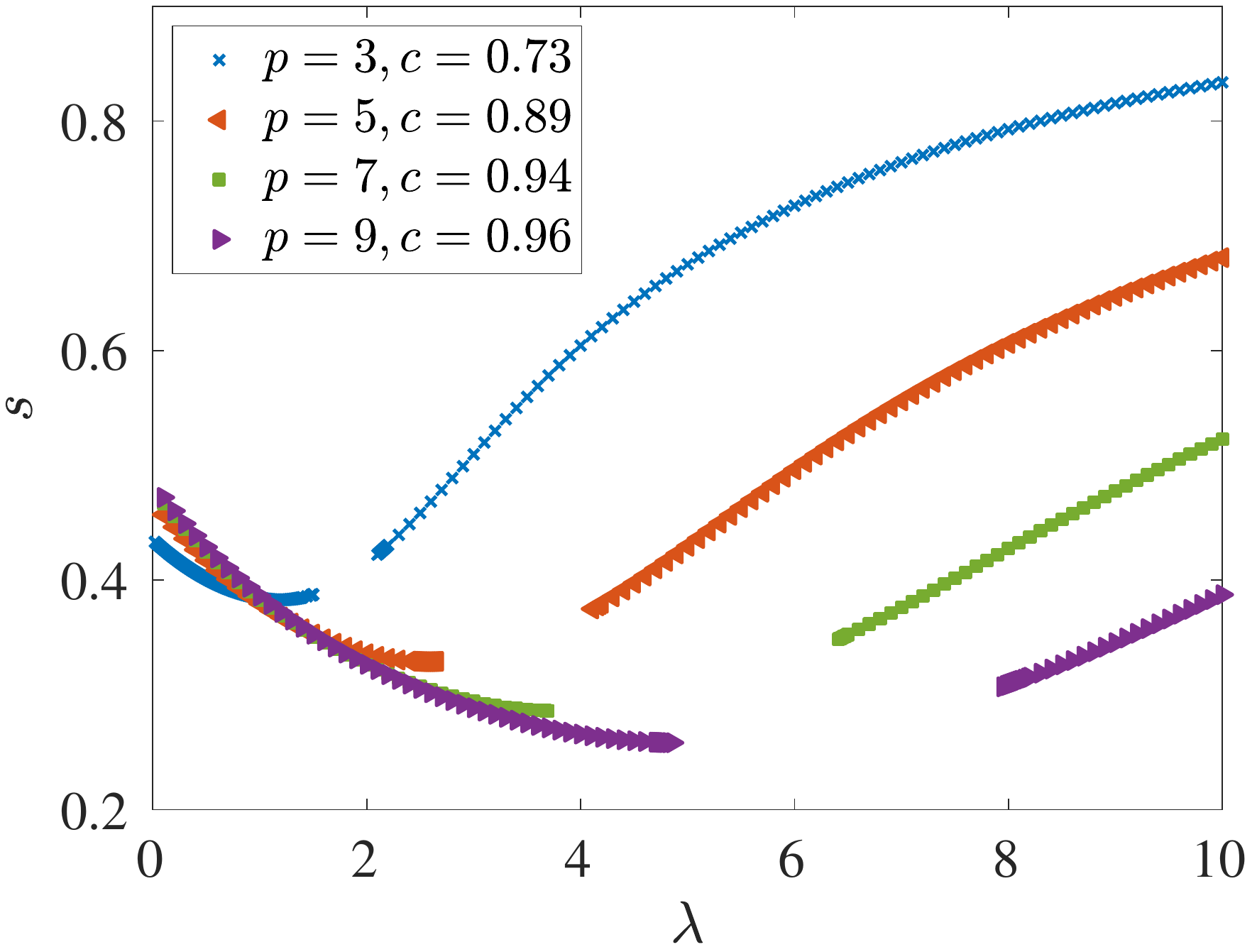}   
   \caption{Phase diagram from mean field theory showing first-order phase transition line for different $p$. Shown here are values of $c$ where there is a range of $\lambda$ values where the first-order phase transition is absent.}  \label{fig:PhaseTransition}
\end{figure}

 However, as the catalyst strength is further increased, the first-order phase transition reemerges because the catalyst biases the wrong configuration and hinders the system from reaching the fully-ferromagnetic ground state (for $c<1$, there is always a subset of qubits that are biased in the `wrong' direction).  Therefore, for sufficiently high $c$, whose value depends on $p$, there is a range of $\lambda$ for which the first-order phase transition can be avoided. We show snapshots of the phase diagram in Fig.~\ref{fig:PhaseTransition}. 
 
%

\subsection{Energy gap behavior}
We supplement our mean-field analysis by calculating the minimum gap $\Delta_\mathrm{min}(\lambda) = \min_{s \in [0,1]} \Delta_{\lambda}(s)$ of the Hamiltonian (Eq.~\eqref{eqt:H_DC}) along the interpolation in the thermodynamic limit using the method developed in Ref.~\cite{Tak2020}. The method calculates the energy gap in the thermodynamic limit by considering fluctuations around the single global minimum of the semi-classical (large-spin) Hamiltonian density:
\bea
\mathcal{H}(s) &=& -(1-s) \left( (1-c) m_2^x + c m_1^x \right) \nonumber \\
&& - \lambda s (1-s) \left( c m_1^z - (1-c) m_2^z \right) \nonumber \\
&&- s \left( c m_1^z + (1-c) m_2^z \right)^p \ ,
\eea
where $m_1^\alpha, m_2^\alpha \in [-1,1]$ correspond to the magnetization density in the $\alpha = x,z$ direction of the two clusters of spins. The method is suited to identify the position of the minimum gap in the parameter regime of $(c,\lambda)$ where the first-order phase transition is absent. We show that the method accurately predicts the thermodynamic limit behavior in Appendix \ref{app:Convergence}. Our calculations of the minimum gap corroborate the conclusions from the mean-field analysis, in that we find that the minimum gap approaches a constant and does not close exponentially for sufficiently large $c$ and $\lambda$ values within the appropriate range.  

However, we also observe a feature that can be expected from our mean-field analysis : the minimum gap exhibits a maximum value as a function of $\lambda$ within the allowed range.  The presence of a maximum can be expected because of the competing effects of increasing $\lambda$: while it biases the energy landscape towards the ferromangetic ground state, it ultimately does energetically favor different spin configurations. We define $\lambda^\ast$ to be the value of $\lambda$ that maximizes the minimum gap, $\lambda^\ast = \arg \max_\lambda \Delta_{\mathrm{min}}(\lambda)$, and the maximum minimum gap is then given by  $\Delta_{\mathrm{min}}^\ast = \Delta_{\mathrm{min}}(\lambda^\ast) = \max_\lambda \Delta_{\mathrm{min}}(\lambda)$.

\begin{figure}[tbp] 
   \centering
   {\includegraphics[width=0.75\columnwidth]{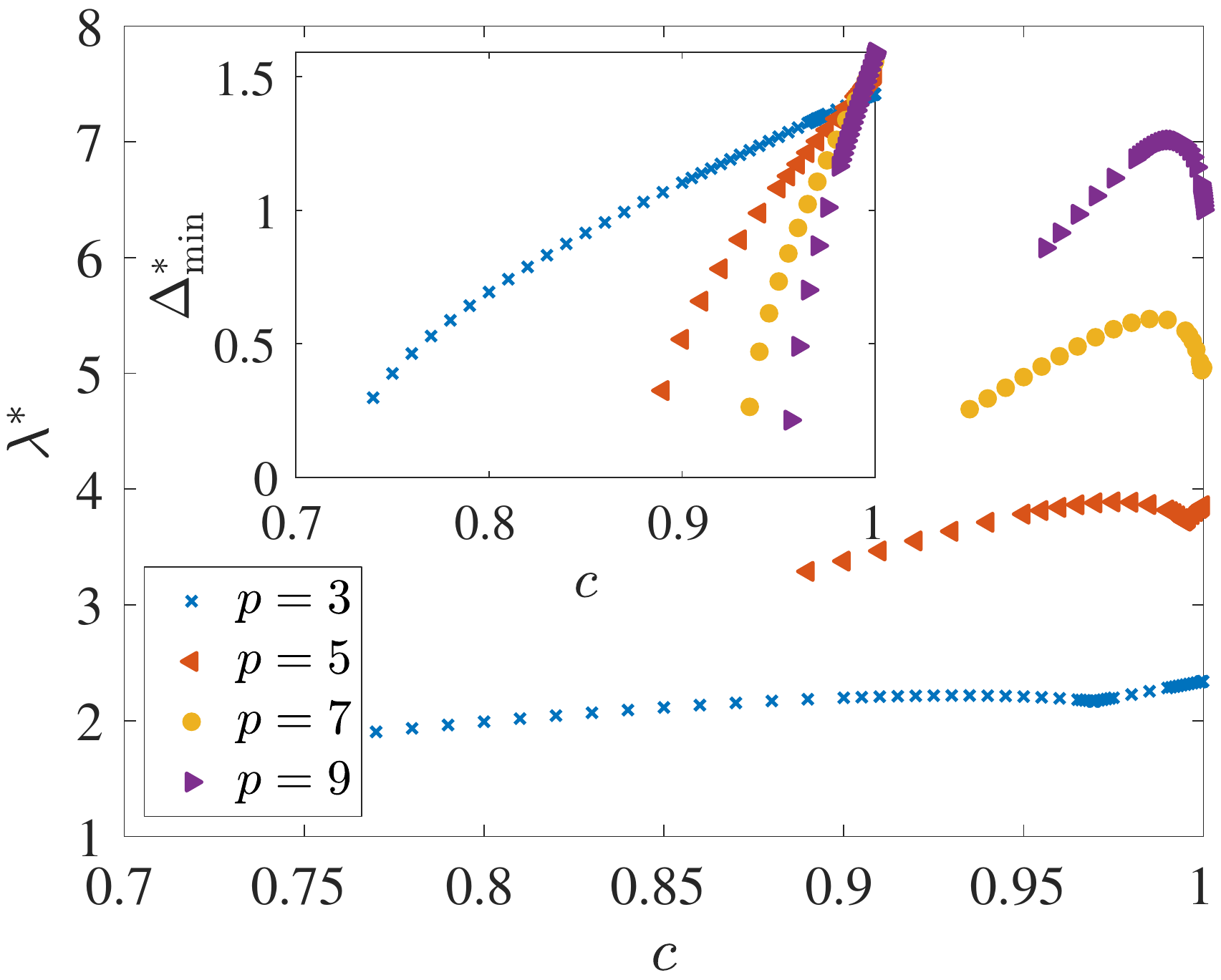}}
   \caption{Behavior of the catalyst strength $\lambda^\ast$, which maximizes the minimum gap, as a function of $c$ for fixed $p$ values. Inset: Behavior of the minimum gap evaluated at  $\lambda^\ast$, $\Delta_{\mathrm{min}}^\ast$, as a function of $c$ for the same $p$ values.}
   \label{fig:FixedpGap}
\end{figure}

While $\Delta_{\mathrm{min}}^\ast$ grows monotonically as $c$ approaches $1$, the behavior of $\lambda^\ast$ is not monotonic with $c$ (see Fig.~\ref{fig:FixedpGap}): it increases, decreases, and increases again (this last region becomes smaller as $p$ increases). We provide an explanation for this behavior in Appendix~\ref{App:NonMonotonicLambda}.

\section{Large $p$ limit: Unstructured Search} \label{sec:LargeP}

So far, we have considered the case of fixed $p$ while take the thermodynamic limit.  We have observed that increasing $p$ in this limit requires increasing $c$ and $\lambda$ to avoid the first-order phase transition (and exponentially closing gap).  We therefore expect that as $p \to \infty$, we must have that $c \to 1$ and $\lambda \to \infty$ in order to avoid the first-order transition.

In order to confirm this expectation, we write our $p$-spin Hamiltonian as:
\bea
H_{\mathrm{P}} &=& -n \left( \ident - \frac{1}{n} \sum_{i=1}^n \left( \ident - \sigma_i^z \right) \right)^p = -n \left( \ident -\frac{2 \text{HW}}{n} \right)^p \nonumber \\
&=& -n \ketbra{0}{0} + n (-1)^p \ketbra{2^n-1}{2^n-1} \nonumber \\
&& - n  \sum_{k=1}^{n-1}  \left(1 - \frac{2 k }{n} \right)^p \sum_{x : |x| =k}\ketbra{x}{x} \ ,
\eea
where $\ket{x}$ denotes the computational basis state with a bit-configuration representing the integer $x$, and $\mathrm{HW} = \frac{1}{2} \sum_{i=1}^n \left( \ident - \sigma_i^z \right)$ is the Hamming weight operator. If we now take the $p \gg 1$, we have that our $p$-spin Hamiltonian is approximately given by
\beq \label{eqt:InfiniteP}
H_{\mathrm{P}} \approx  -n \ketbra{0}{0} - n (-1)^p \ketbra{2^n-1}{2^n-1} \ .
\eeq
This Hamiltonian is effectively the oracle Hamiltonian of unstructured search \cite{Roland:2002ul,Jorg:2010qa}, except for the factor of $n$ and the additional penalty on the conjugate of the marked state.  (Here $\ket{0}$ is the marked state, and $\ket{2^n-1}$ is its conjugate.) The factor of $n$ arises because we had made our $p$-spin Hamiltonian extensive (see Eq.~\eqref{eqt:pSpin}), and it leads to a gap between the ground state and the (degenerate) first excited state that grows with $n$.  

\begin{figure}[htbp] 
   \centering
   {\includegraphics[width=0.75\columnwidth]{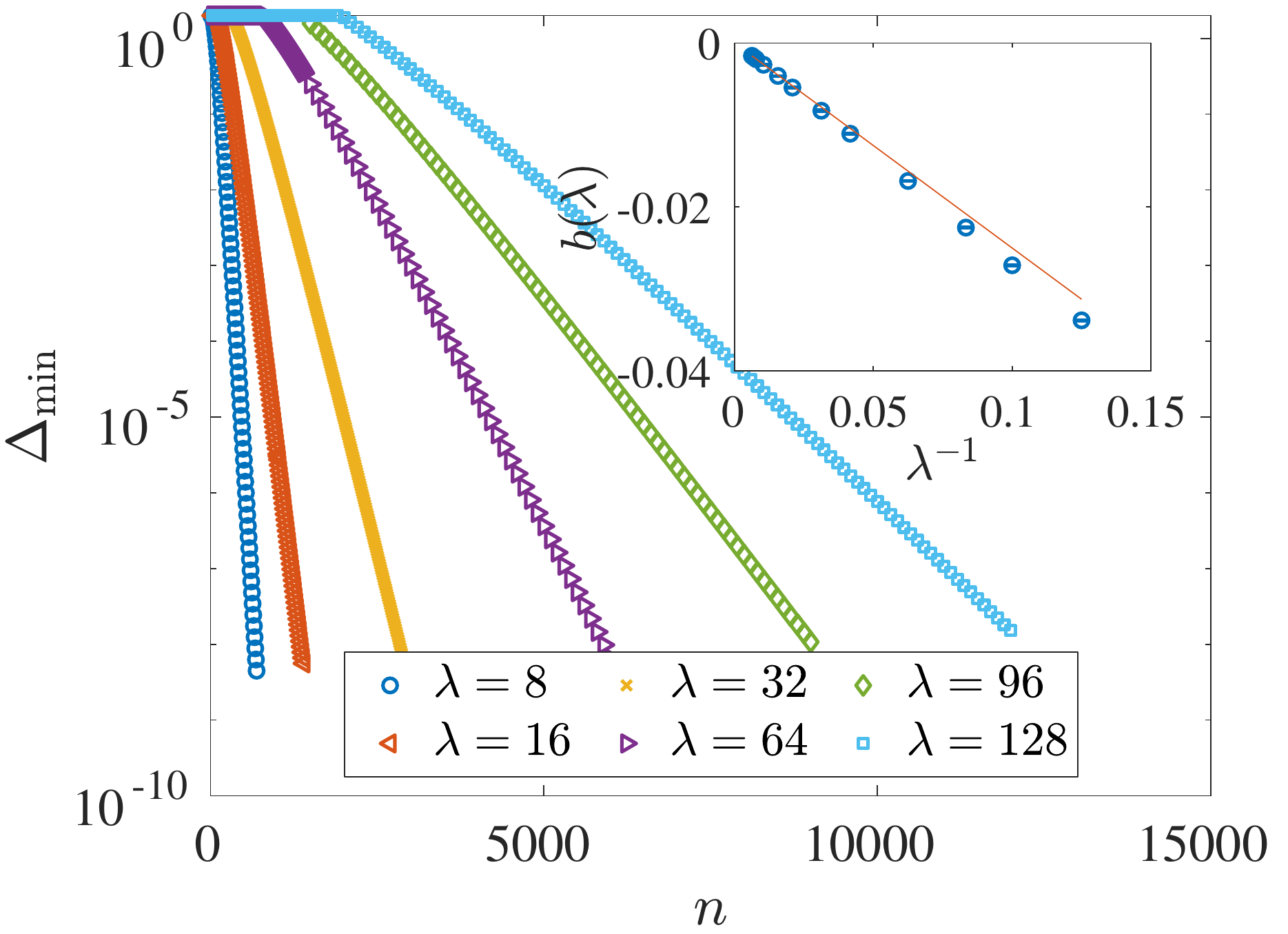}}
   \caption{Behavior of the minimum gap $\Delta_{\mathrm{min}}$ at $c=1$ for the $p\gg1$ problem Hamiltonian in Eq.~\eqref{eqt:InfiniteP}. Inset: The fit of the exponential scaling parameter of the minimum gap, $\Delta_{\mathrm{min}} \sim \exp( b(\lambda) n)$, with $b(\lambda)$ approaching $1/(2\lambda)$ as $\lambda$ becomes large, as indicated by the red line.}
   \label{fig:LargePGap}
\end{figure}

We focus on the case of $p$ odd.  Even for $c=1$, when the bias is entirely on the final ground state, for a fixed $\lambda$ there is always a system size above which the gap scaling is exponential (see Fig.~\ref{fig:LargePGap}). Thus for fixed $\lambda$, our bias Hamiltonian is not able to eliminate the first-order phase transition typically associated with unstructured search. Above this size, the minimum gap scales as $\Delta_{\mathrm{min}} \sim \exp( n /(2\lambda))$, so it follows that in order to ensure a constant gap as the system size scales, $\lambda$ must scale linearly with the system size.  This finding is then consistent with our expectation from our fixed $p$ analysis of the previous section.  A similar result was also found in Ref.~\cite{Bri2019}.

\section{Weak-strong cluster problem} \label{sec:WeakStrong}
\begin{figure}[t] 
  \centering
  \includegraphics[width=0.5\columnwidth]{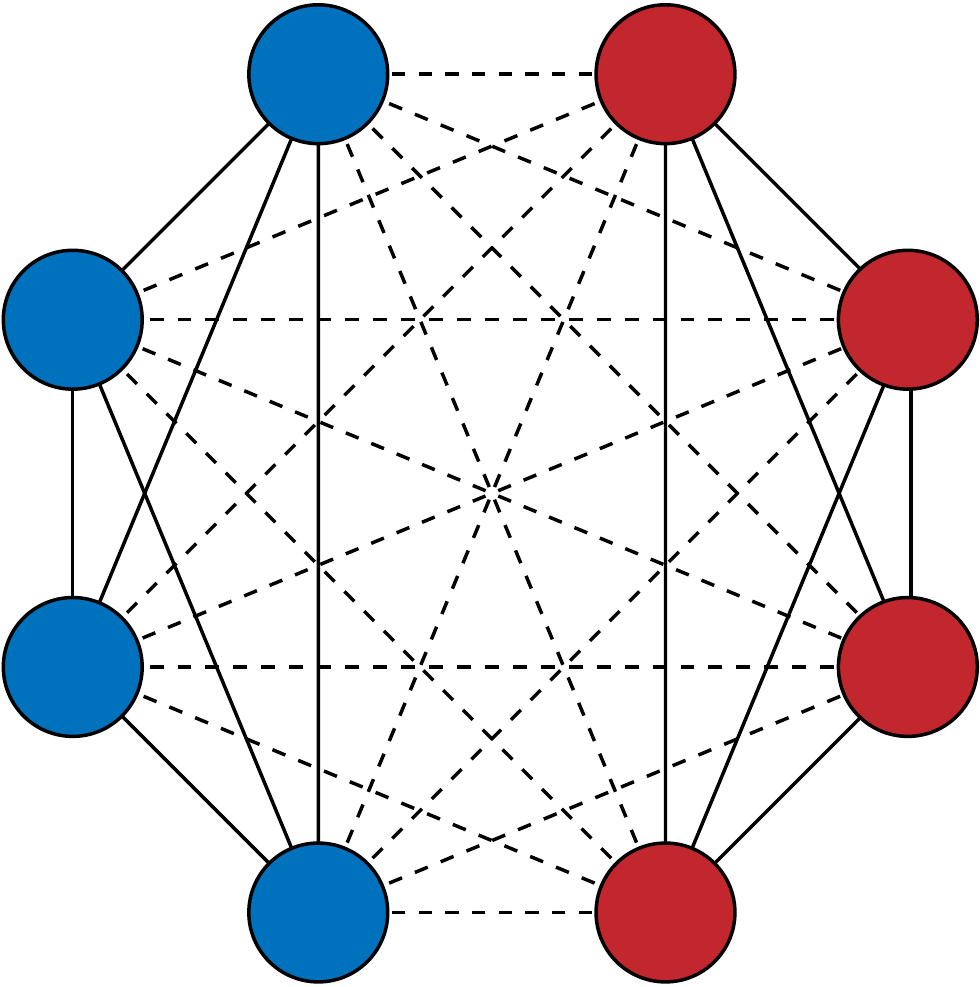} 
  \caption{Weak-Strong Cluster problem.  Blue (left) and red (right) circles correspond to the two clusters of spins, one with a strong local field and the other with a weak local field in the opposite direction.  Solid edges correspond to intra-cluster ferromagnetic couplings, and dashed edges correspond to inter-cluster ferromagnetic couplings.}
  \label{fig:StrongWeakCluster}
\end{figure}

Our next example is the prototypical large-spin tunneling problem studied in Ref. \cite{GoogleTunneling,nonStoq,Tak2020}. The problem is characterized by two fully-connected clusters of spins, one with a `strong' local field in one direction, and the other with a `weak' local field pointing in the opposite direction, depicted in Fig.~\ref{fig:StrongWeakCluster}.  We will denote the former as the strong cluster and the latter as the weak cluster. In the standard QA protocol, the interpolation exhibits an exponentially closing gap, associated with the global minimum changing from the weak cluster being anti-aligned with the strong cluster to being aligned with the strong cluster. This event is then associated with the tunneling of $\mathcal{O}(n/2)$ spins. 

The role of the diagonal catalyst can be readily understood in this case.  
We consider the following interpolating Hamiltonian:
\begin{eqnarray} \label{eqt:H2}
&&H(s) = -2 (1-s) \left(S_1^x + S_2^x \right)  - s \left(  2 h_1 S_1^z - 2 h_2 S_2^z   \right. \nonumber \\
&& \left. +\frac{4}{n} \left((S_1^z)^2 +(S_2^z)^2 + S_1^z S_2^z\right) \right)   - \lambda s(1-s) \sum_{i=1}^n \epsilon_i \sigma_i^z \ , 
\end{eqnarray}
where $S_1^\alpha = \frac{1}{2} \sum_{i=1}^{n/2} \sigma_i^\alpha$, $S_2^\alpha = \frac{1}{2} \sum_{i=n/2 + 1}^n \sigma_i^\alpha$.  For simplicity, we restrict to the case where $n$ is even.  We take $h_1 = 1$ and $h_2 = 0.49$, such that the ground state at $s=1$ has eigenvalues  $(+1,+1)$ under $\frac{2}{n} S_1^z$ and $\frac{2}{n} S_2^z$ respectively. The terms associated with the local fields on the weak cluster (the terms proportional to $h_2$) are violated by the ground state at $s=1$, but the instantaneous ground during the interpolation initially aligns with these local fields, which leads to the exponentially closing gap.  A diagonal catalyst that reduces the alignment with the weak cluster local field during the interpolation can thus be expected to eliminate the exponentially closing gap. This intuitive argument already suggests that if we consider a catalyst of the form:
\beq
H_\mathrm{B} = - \sum_{i=1}^{n/2} \sigma_i^z - \sum_{i=n/2+1}^{\frac{n}{2}( 1 + c)} \sigma_i^z  +  \sum_{i= \frac{n}{2}( 1 + c)+1}^{n} \sigma_i^z \ ,
\eeq
where we correctly bias the strong cluster and $\frac{n}{2} c$ qubits of the weak cluster and incorrectly bias $\frac{n}{2} (1-c)$ qubits of the weak cluster, then we should be able to eliminate the exponentially closing gap  for $c > 1/2$.

\begin{figure}[t] 
  \centering
 \includegraphics[width=0.75\columnwidth]{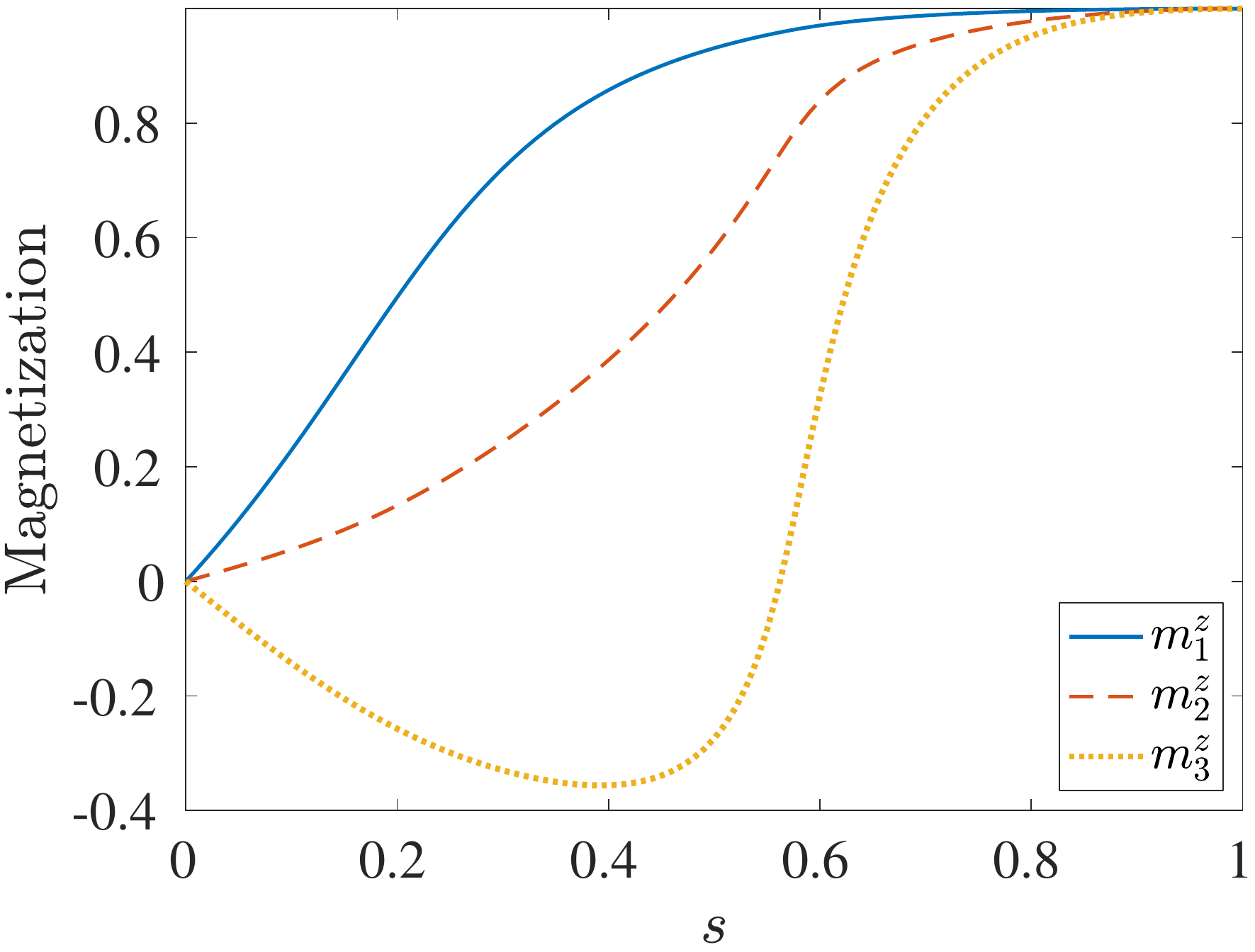}   
  \caption{Magnetization at the global minimum of the semiclassical Hamiltonian for the Weak-Strong Cluster problem (Eq.~\eqref{eqt:WSH}) for the case of $c=0.6$ and $\lambda=1$ for which the first-order phase transition is avoided.}\label{fig:StrongWeakClusterMagnetization2}
\end{figure}

In order to validate this intuition, we first consider the free energy density in the zero-temperature limit as before:
\bea
f(m_1,m_2) &=& \frac{s}{4} \left( m_1^2 + m_2^2 + m_1 m_2 \right)  \nonumber \\
&& \hspace{-2cm} -\frac{1}{2} \sqrt{s^2\left( m_1 + \frac{m_2}{2} + h_1 + \lambda(1-s)\right)^2 + (1-s)^2} \nonumber \\
&& \hspace{-2cm}-\frac{c}{2} \sqrt{s^2\left( m_2 + \frac{m_1}{2} -h_2 + \lambda(1-s)\right)^2 + (1-s)^2} \nonumber \\
&& \hspace{-2cm} -\frac{1-c}{2} \sqrt{s^2\left( m_2 + \frac{m_1}{2} -h_2 - \lambda(1-s)\right)^2 + (1-s)^2} \ . \nonumber \\
\eea
For a sufficiently large $c$ above $1/2$, we find we can avoid any discontinuous jump in the free energy for a sufficiently large $\lambda$. The mean field analysis can be corroborated by calculating the gap in the thermodynamic limit. The semi-classical Hamiltonian density is given by:%
\bea \label{eqt:WSH}
\mathcal{H}(s) &=& -\frac{1}{2}(1-s)\left( m_1^x + c m_2^x + (1-c) m_3^x \right) \nonumber \\
&& - \frac{\lambda}{2} s (1-s) \left( m_1^z + c m_2^z - (1-c) m_3^z \right) \nonumber \\
&& - s \left[ \frac{h_1}{2} m_1^z - \frac{h_2}{2} \left( c m_2^z + (1-c) m_3^z \right)  \right. \nonumber \\
&&  +  \frac{1}{4} (m_1^z)^2 + \frac{1}{4} \left( c m_2^z + (1-c) m_3^z \right)^2 \nonumber \\
&& \left. + \frac{1}{4} m_1^z \left( c m_2^z + (1-c) m_3^z \right)  \right] \ ,
\eea
where $m_1^\alpha$, $\alpha = x,y,z$, is the magnetization of the strong cluster, $m_2$ is the magnetization of the $c$-fraction of spins in the weak cluster with the correct bias, and $m_3$ is the magnetization of the $(1-c)$-fraction of spins in the weak cluster with the incorrect bias. We show in Fig.~\ref{fig:StrongWeakClusterMagnetization2} how these three magnetizations behave in the absence of a closing gap, and the behavior is consistent with our mean field analysis.
\section{Bottlenecks introduced by diagonal catalysts} \label{sec:Bottleneck}
%
Our study of the $p$-spin and the Weak-Strong Cluster problems has revealed under what conditions we can expect a diagonal catalyst to enhance the performance of the adiabatic algorithm. Crucial to this success has been applying an energetic bias on states that are close in Hamming distance to the desired ground state. However, this implicitly assumes that we have a good reason (without knowing the ground state) to apply a bias on these states as opposed to other states.  In the case of the $p$-spin model, states with low Hamming distance are also states with low energy, so biasing these states is a reasonable procedure.

However, it is easy to construct instances where this is not the case, meaning the low energy states are actually very far in Hamming distance from the true ground state.  In this case, biasing towards low energy states can have a severely detrimental effect on the performance of the algorithm.  We demonstrate this next with some striking cases.  Our constructions are based on `perturbative crossings' \cite{PhysRevA.80.062326}, a known bottleneck of QAO.

\subsection{Perturbative Crossings}
\begin{figure}[t] 
   \centering
   \includegraphics[width=0.9 \columnwidth]{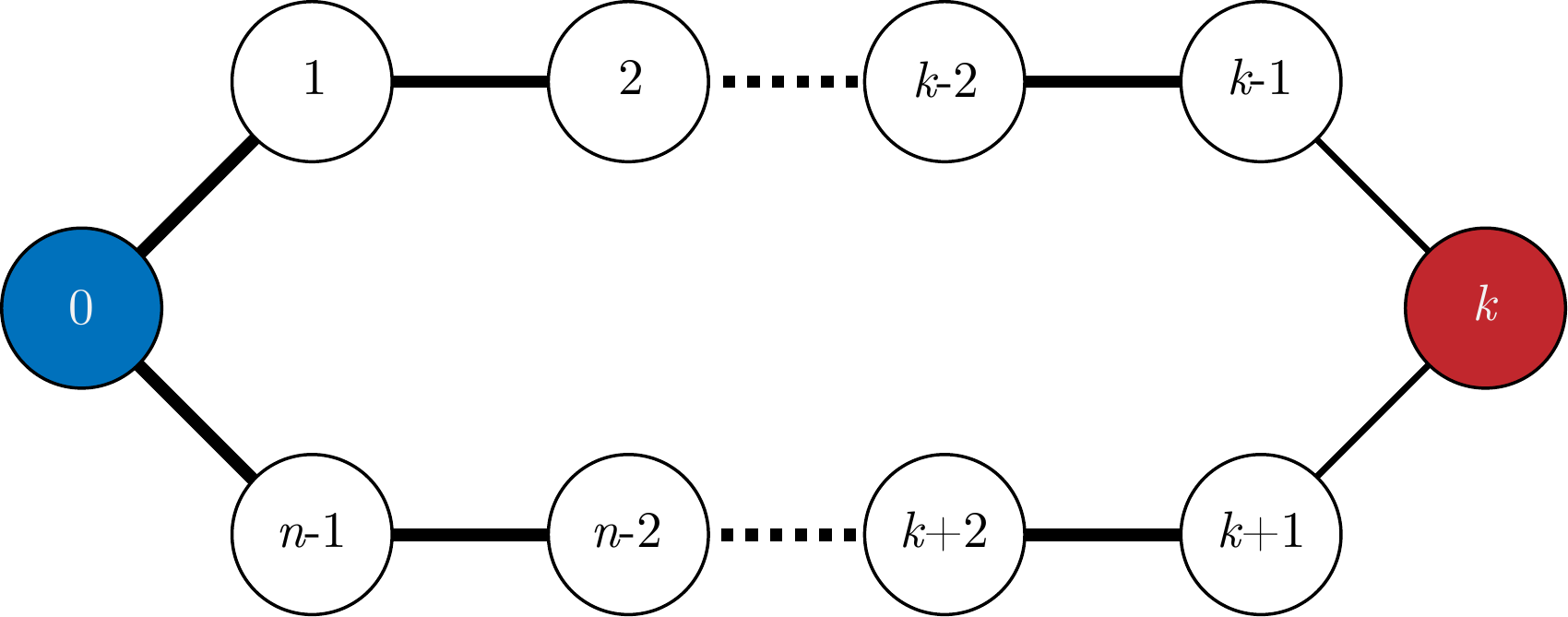} 
   \caption{Illustration of the Ising instance described by Eq.~\eqref{eqt:LoopGadget}.  The numbers inside the circles denote the labeling of the qubits.  The links between circles denote ferromagnetic couplings (note the thickness denotes the strength of the coupling).  The red circle (left-most) and blue circle (right most) denote qubits with non-zero Ising local fields in opposite directions.}
   \label{fig:LoopGadget}
\end{figure}

We first consider a one-dimensional Ising Hamiltonian on $n = 2 k$ spins given by:
\beq \label{eqt:LoopGadget}
H_{\mathrm{P}}  = \frac{1}{R} \left( (R-1) \sigma_0^z - R \sigma^z_{k} -\sum_{i=0}^{n-1} J_{i,i+1} \sigma_i^z \sigma_{i+1}^z \right)
\eeq
with  $R \geq 4$, $\sigma_n^z \equiv \sigma_0^z$ (periodic boundary conditions), and
 \beq
 J_{i,i+1} = \left\{\begin{array}{rl}
 R/2 & \mathrm{if} \ i = k-1, k \\
 R & \mathrm{otherwise}
 \end{array} \right.
 \eeq 
This system, depicted in Fig.~\ref{fig:LoopGadget}, has anti-aligned local fields on opposite ends of the periodic chain with  magnitudes $1$ and $(1-1/R)$ respectively and Ising couplings that are uniform everywhere except at one end of the chain.  
The ground state is given by the all-zero bit string ($\ket{\phi} \equiv \ket{0_{n-1} 0_{n-2} \dots 0_{0}}$), where the subscripts denote the qubit index as labeled in Fig.~\ref{fig:LoopGadget}, with energy $E_0$, and the first excited state is doubly degenerate with states corresponding to the all-one state ($\ket{\psi} \equiv \ket{1_{n-1} 1_{n-2} \dots 1_{0} }$) and the same state except with a 0 on the $k$-indexed qubit ($\ket{\eta} \equiv \ket{1_{n-1} 1_{n-2} \dots 1_{k+1} 0_k 1_{k-1} \dots 1_0}$) with energy $E_1$.

Let us now consider the annealing protocol with a diagonal catalyst, Eq.~\eqref{eqt:H_DC}. We perform a perturbative analysis away from the point $s = 1$, with the perturbation parameter being $\Gamma = 1-s$. The Hamiltonian up to first order is given by:
\beq \label{eqt:HfirstOrder}
H^{(1)}(\Gamma) = H_{\mathrm{P}} + \Gamma V_1 = H_{\mathrm{P}} + \Gamma \left( H_{\mathrm{D}} + H_{\mathrm{I}} - H_{\mathrm{P}} \right) \ . 
\eeq
At this order in perturbation theory, the first excited states $\ket{\psi}, \ket{\eta}$ and the ground state $\ket{\phi}$ are not coupled, i.e. $\bra{\psi}V_1 \ket{\phi} = \bra{\eta} V_1 \ket{\phi} = 0$.  Thus, in order to determine how the first excited state degeneracy is broken, we can restrict ourselves to the degenerate subspace of $\ket{\psi}, \ket{\eta}$ at this order.  In this subspace, we can represent $V_1$ as:
\beq
V_1 \rightarrow \left( \begin{array}{cc}
-E_1 + \varepsilon_1 - \lambda \epsilon_{k} & 1 \\
1 & -E_1 + \varepsilon_1 + \lambda \epsilon_{k} 
\end{array} \right)_{\ket{\psi}, \ket{\eta}} \ ,
\eeq
where $\varepsilon_1 =  \lambda \sum_{i \neq k} \epsilon_i$. Specifically, the degeneracy is not only broken by the transverse field but also by the diagonal catalyst.  Thus at first order in perturbation theory, the (non-degenerate) instantaneous first excited state energy is given by $E_1 - \Gamma(E_1 - \varepsilon_1 +  \sqrt{1 + \lambda^2})$.  The instantaneous ground state energy is given by $E_0 - \Gamma (E_0 + \varepsilon_0)$, where $\varepsilon_0 = \lambda \sum_i   \epsilon_i$.  Therefore, the instantaneous gap at first order in perturbation theory is given by:
\beq
\Delta = \Delta_0 - \Gamma \left(\Delta_0 + \sqrt{ 1 + \lambda^2 }  - \varepsilon_1 - \varepsilon_0) \right) \ ,
\eeq
where $\Delta_0 = E_1 - E_0$. Note that $\varepsilon_1 + \varepsilon_0 = 2 \lambda \sum_{i \neq k} \epsilon_i + \lambda \epsilon_{k}$. The perturbative crossing is then expected to occur when this gap becomes zero, which occurs at a value of $\Gamma_{\mathrm{DC}}^* = \Delta_0 /  \left(\Delta_0 + \sqrt{ 1 + \lambda^2 }  - \varepsilon_1 - \varepsilon_0) \right)$. 

If we are to use a diagonal catalyst that energetically favors the first excited state $\ket{\psi}$, we would have $ - \varepsilon_1 - \varepsilon_0 = (2n - 1)\lambda $, which means that the perturbative crossing approaches $s = 1$ as the problem size grows, $\Gamma_{\mathrm{DC}}^* \sim 1/n$.
Since the gap at the perturbative crossing scales as $\Gamma_\ast^n$, our results suggests a factorial scaling for the gap, $\Delta_{\mathrm{min}} \sim n^{-n}$.

This is to be contrasted with the standard protocol, where the instantaneous gap at first order would be given by:
\beq
\Delta_{\mathrm{S}}= \Delta_0 - \Gamma \left(\Delta_0 + 1 \right) \ ,
\eeq
and the perturbative crossing occurs at $\Gamma_{\mathrm{S}}^* = \Delta_0 / (\Delta_0 + 1)$, which is independent of problem size.  Thus we only expect an exponential scaling for the gap in this case, $\Delta_{\mathrm{min}} \sim c^{-n}$.

\begin{figure}[t] 
   \centering
   \includegraphics[width=0.75 \columnwidth]{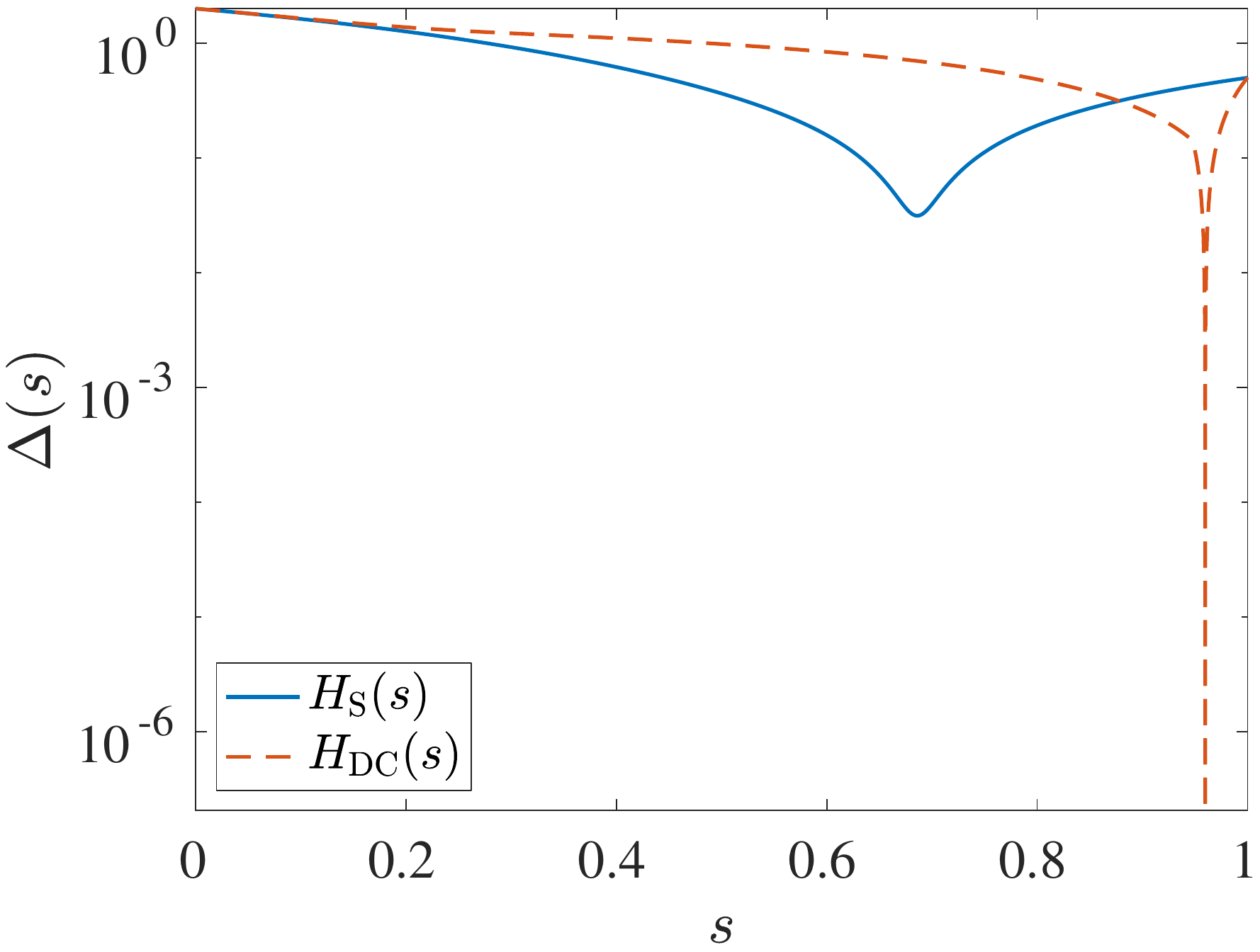} 
   \caption{Energy gap $\Delta(s)$ for $n=6$ and $R=4$ for the problem Hamiltonian in Eq.~\eqref{eqt:LoopGadget} using the standard protocol $H_{\mathrm{S}}(s)$ and the one with a diagonal catalyst $H_{\mathrm{DC}}(s)$ with $H_{\mathrm{I}} = \sum_{i=0}^{n-1} \sigma_i^z$ and $\lambda = 1$.  This amounts to taking $\varepsilon_1 = - \lambda (n-1) \ , \varepsilon_0 = -\lambda n$.  For $H_{\mathrm{DC}}$, first order perturbation theory predicts a level crossing at $s^\ast_{\mathrm{DC}} = 1 - \Gamma^\ast_{\mathrm{DC}} \approx 0.961283$, whereas the true avoided crossing calculated numerically occurs at $s_\mathrm{min} \approx 0.959768$.  For $H_{\mathrm{S}}$, first order perturbation theory predicts a level crossing at $s^\ast_{\mathrm{S}} = 2/3$, whereas the true avoided crossing calculated numerically occurs at $s_\mathrm{min} \approx 0.707325$.}
   \label{fig:GapLoopGadget}
\end{figure}

To illustrate how this can affect performance, we show the energy gap of $H_{\mathrm{S}}(s)$ and $H_{\mathrm{DC}}(s)$ with $H_{\mathrm{B}} = \sum_{i=0}^{n-1} \sigma_i^z$ in Fig.~\ref{fig:GapLoopGadget}, where we have chosen to bias towards the state $\ket{\psi}$.  The minimum gap using the diagonal catalyst Hamiltonian is almost 5 orders of magnitude smaller than the standard Hamiltonian even for a 6 qubit problem.
\subsection{Inducing a Perturbative Crossing}
%
In the previous example, both $H_{\mathrm{S}}(s)$ and $H_{\mathrm{DC}}(s)$ exhibit a perturbative crossing.  We now consider an example where only $H_{\mathrm{DC}}(s)$ exhibits a perturbative crossing.  We consider a modification of our previous Hamiltonian such that it now takes the form
\beq \label{eqt:LoopGadget2}
H_{\mathrm{P}}  = \frac{1}{R} \left( R \sigma_0^z - (R-1) \sigma^z_{k} -\sum_{i=0}^{n-1} J_{i,i+1} \sigma_i^z \sigma_{i+1}^z \right) \ .
\eeq
For this Hamiltonian, the ground state is the all-one state ($\ket{\psi}$), and the doubly degenerate first excited is now given by the all-zero state ($\ket{\phi}$) and the state $\ket{\eta}$.  

In the standard annealing protocol, this problem does \emph{not} exhibit a perturbative crossing.  However, in the presence of a diagonal catalyst, a perturbative crossing can be induced.  For example, if we take $H_{\mathrm{I}} = - \sum_{i} \sigma_i^z$, such that the catalyst biases  the state $\ket{\phi}$, then in the first excited state subspace, the first order correction to the Hamiltonian in the perturbative parameter $\Gamma$ is given by
 \beq
V_1 \rightarrow \left( \begin{array}{cc}
-E_1 - n & 0 \\
0 & -E_1 + n - 1
\end{array} \right)_{\ket{\phi},\ket{\eta}} \ .
\eeq
The instantaneous energy gap at first order in perturbation theory is then given by:
\beq
\Delta_{\mathrm{DC}}^{(1)} = \Delta_0 - \Gamma (\Delta_0 + 2 n) \ ,
\eeq
and the perturbative crossing is expected to occur at $\Gamma_{\mathrm{DQ}}^* = \Delta_0 /  \left(\Delta_0 + 2n) \right)$.  We thus can expect the minimum gap to scale factorially in the system size.  

We provide another example of such an induced perturbative crossing in Appendix~\ref{app:LoopGadget3}.
\section{Discussion and Conclusions} \label{sec:Conclusions}

We have illustrated how introducing a catalyst Hamiltonian that is diagonal in the computational basis to the standard interpolating Hamiltonian of QAO can exhibit an exponential improvement in the efficiency of the algorithm for solving the ferromagnetic $p$-spin model and the weak-strong cluster problem.  Our results are analogous to the exponential improvement observed for ARA \cite{PhysRevA.98.022314}, and the diagonal catalyst approach highlights the mechanism for this improvement: the Hamiltonian $H_{\mathrm{B}}$ biases the energy landscape towards the solution, and if the bias is sufficiently strong it can eliminate discontinuous jumps in the global minimum of the landscape, which are the bottleneck of the algorithm.  

Our work also highlights the danger of over-selling this approach in the broader context of solving for the ground state of hard optimization problems using QAO. In the ferromagnetic $p$-spin models, the energy landscape is simple: the Hamming distance of a state from the ground state is directly correlated with its energy, so a bias Hamiltonian that progressively favors lower Hamming weight states will naturally help the algorithm reach the target ground state. However, in general, we expect there to be no correlation between a state's Hamming distance and its closeness in energy to the ground state, and in this more general case there is no evident prescription of how to pick the bias Hamiltonian. We highlighted this using examples where biasing low energy but high Hamming distance states impedes the algorithm; this has already been observed in the results of Ref.~\cite{Perdomo-Ortiz:2011fh}. 

The operators appearing in the DC and ARA Hamiltonian are the same, and the only difference is how each term is turned on and off during the interpolation.  Because the spectral gap is purely a property of the instantaneous Hamiltonian and not of the interpolation used to reach that Hamiltonian, we can expect both approaches to achieve the same minimum gap, assuming that both approaches are able to enact sufficiently rich interpolations. This suggests that the two approaches should achieve the same adiabatic algorithm efficiency.  From this  point of view, there does not appear to be a strong reason to prefer one approach over the other.  However, there is one advantage of using a catalyst over ARA that we can identify.   In ARA, the Hamiltonian $H_{\mathrm{B}}$ needs to energetically bias a single bit-string in order to ensure there is a unique initial ground state. This bit-string will have a subset of bits that disagree with the ground state of $H_{\mathrm{P}}$.  However, in the case of the catalyst, we need not place a bias on all qubits since the initial state of the algorithm is still the ground state of the driver Hamiltonian $H_\mathrm{D}$.  Thus with a catalyst, we can in principle avoid or mitigate biasing qubits incorrectly by not having $H_\mathrm{B}$ act on all qubits.  While this may favor the catalyst approach, we believe that it is more important to identify which approach is more robust to experimental imperfections and noise.  We hope to address this question in future work. 

\begin{acknowledgments}
Computation for the work described in this paper was supported by the University of Southern California's Center for High-Performance Computing (hpc.usc.edu) and by ARO grant number W911NF1810227.
The research is based upon work (partially) supported by the Office of
the Director of National Intelligence (ODNI), Intelligence Advanced
Research Projects Activity (IARPA) and the Defense Advanced Research Projects Agency (DARPA), via the U.S. Army Research Office
contract W911NF-17-C-0050. The views and conclusions contained herein are
those of the authors and should not be interpreted as necessarily
representing the official policies or endorsements, either expressed or
implied, of the ODNI, IARPA, DARPA, or the U.S. Government. The U.S. Government
is authorized to reproduce and distribute reprints for Governmental
purposes notwithstanding any copyright annotation thereon.
\end{acknowledgments}

%

\appendix 
\section{Deformation of the mean field free energy landscape in the presence of the catalyst} \label{app:Landscape}
\begin{figure}[h] 
   \centering
 \includegraphics[width=0.75\columnwidth]{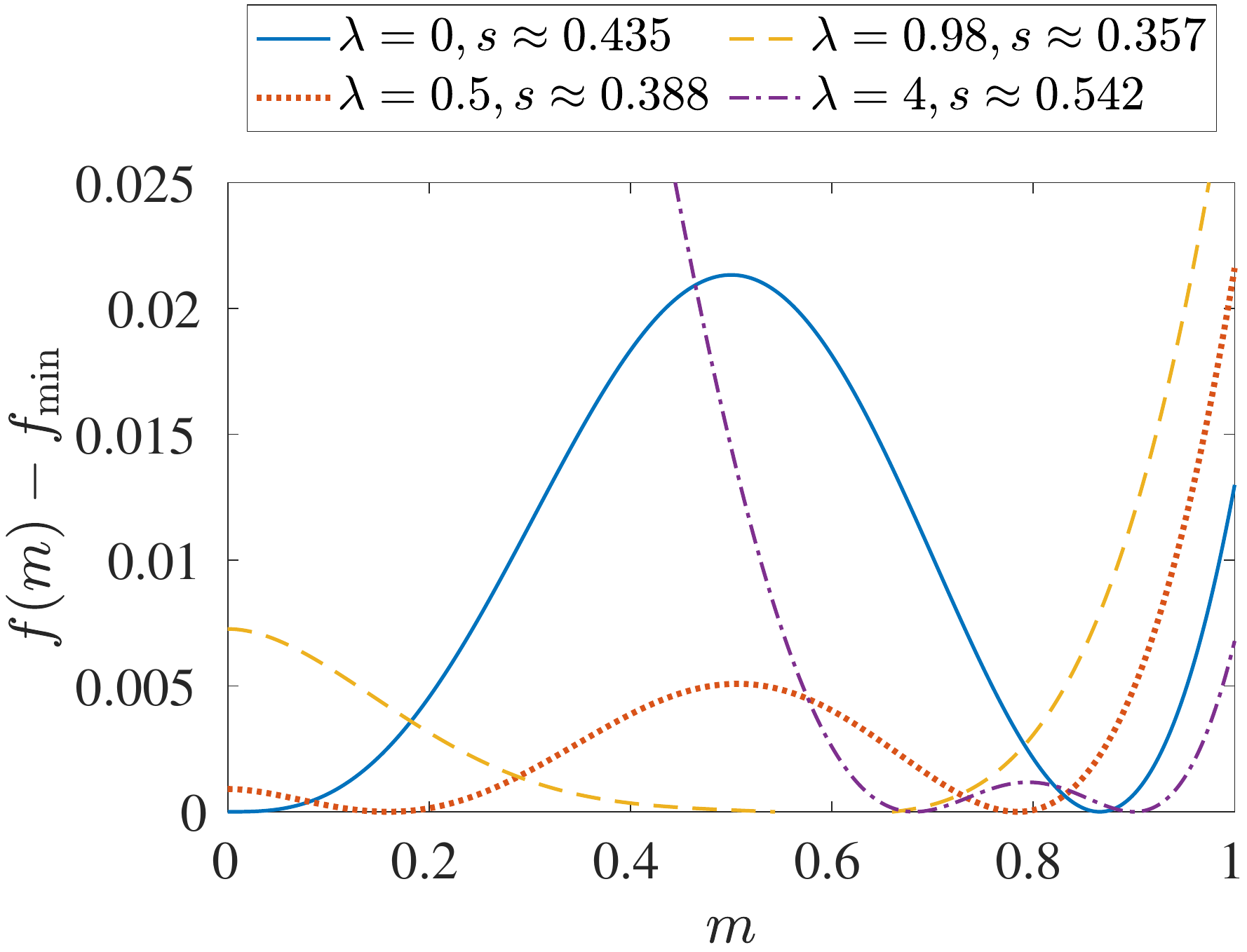} 
   \caption{The mean-field free energy (Eq.~\eqref{eqt:pSpinFreeEnergy}) for the $p$-spin model with $p=3$ and $c=0.8$ and different $\lambda$ values.  The constant $f_{\mathrm{min}}$ is the value of the free energy at the minimum, and it is subtracted off so that all minima occur at zero. For $\lambda = 0$, and $0.5$, the values of $s$ is chosen to correspond to the point along the interpolation where the free energy exhibits doubly-degenerate minima.  For $\lambda = 0.97$, there is only a single minimum along the entire interpolation.
}   \label{fig:DC_FreeEnergy}
\end{figure}

 In the absence of a catalyst, $f$ exhibits a degenerate double-well with one minimum at $m_1=0$ and the other at $m_2$, with $m_2 > m_1$.  When the catalyst is turned on and increased in strength, the first minimum $m_1$ now occurs at a non-zero value, while $m_2$ remains at (approximately) the same position.  There is therefore a reduction in the separation between the two minima.  Above a critical value $c$ and when the catalyst is sufficiently strong, the two minima merge and the double well vanishes, and only a single minimum is realized during the entire anneal.  We depict this behavior in Fig.~\ref{fig:DC_FreeEnergy}. 

\section{Gap in the thermodynamic limit} \label{app:Convergence}
\begin{figure}[t] 
   \centering
 \includegraphics[width=0.75\columnwidth]{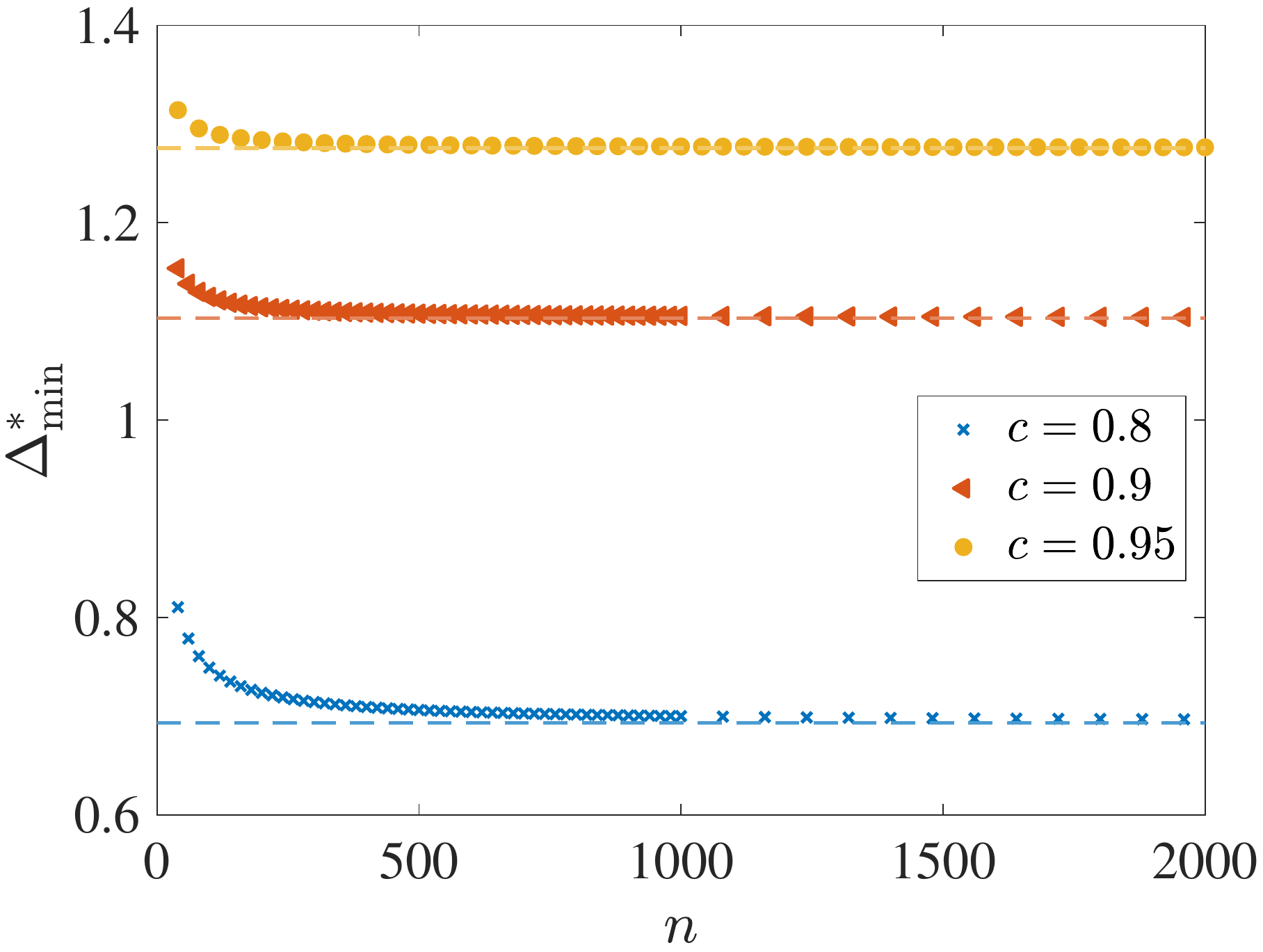}   
   \caption{Convergence of the finite $n$ minimum gap values (data points) to the thermodynamic value (shown as dashed lines) for $p=3$.} \label{fig:AsymptoticGap}
\end{figure}

We show how the the method of Ref.~\cite{PhysRevA.98.022314} accurately predicts the minimum gap in the thermodynamic limit for the models considered in Sec.~\ref{sec:pSpin}. We use the fact that the interpolating Hamiltonians are permutation symmetric for two  subsets of the qubits.  Since the initial ground state obeys these symmetries as well, we can focus on the symmetric subspace to which the ground state belongs.  For the case of the ferromagnetic $p$-spin problem Hamiltonian, this subspace is spanned by the tensor product of Dicke states \cite{Dic1954}, $\left\{ \ket{j_1,m_1} \otimes \ket{j_2,m_2} \right\}$, with $j_1 = c n/2, j_2 = (1-c)n/2$ and $m_1 = -j_1, \dots, j_1, m_2 = -j_2, \dots, j_2$.   This subspace is only $(2 j_1 + 1) \times (2j_2 + 1)$ dimensional.  In Fig.~\ref{fig:AsymptoticGap}, we show how the minimum gap asymptotes to the thermodynamic limit value. 

\section{Non-monotonicity in $\lambda^\ast$} \label{App:NonMonotonicLambda}
\begin{figure}[htbp] 
   \centering
   \subfigure[\ $c=0.9$]{\includegraphics[width=0.75\columnwidth]{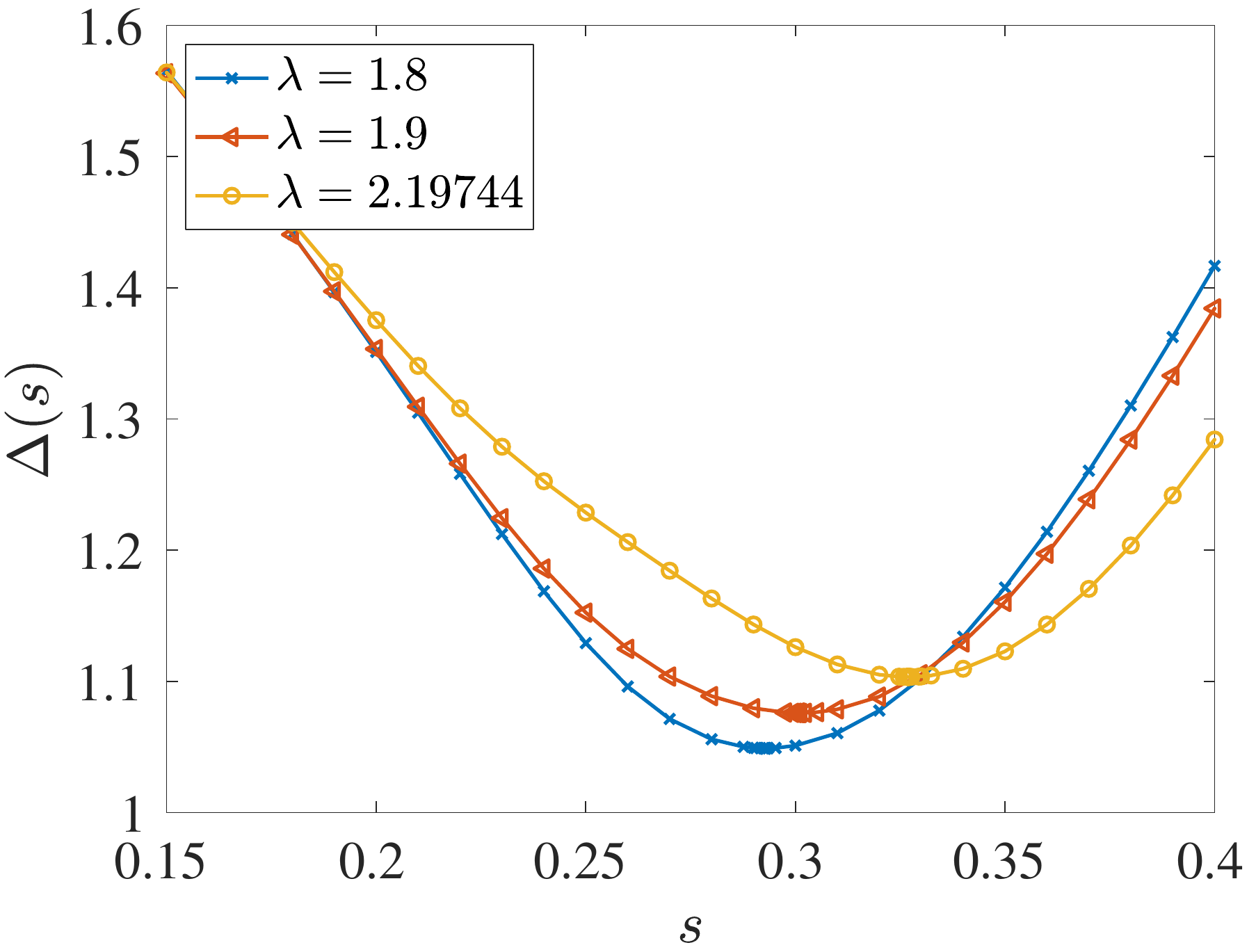}}
   \subfigure[\ $c=0.93$]{\includegraphics[width=0.75\columnwidth]{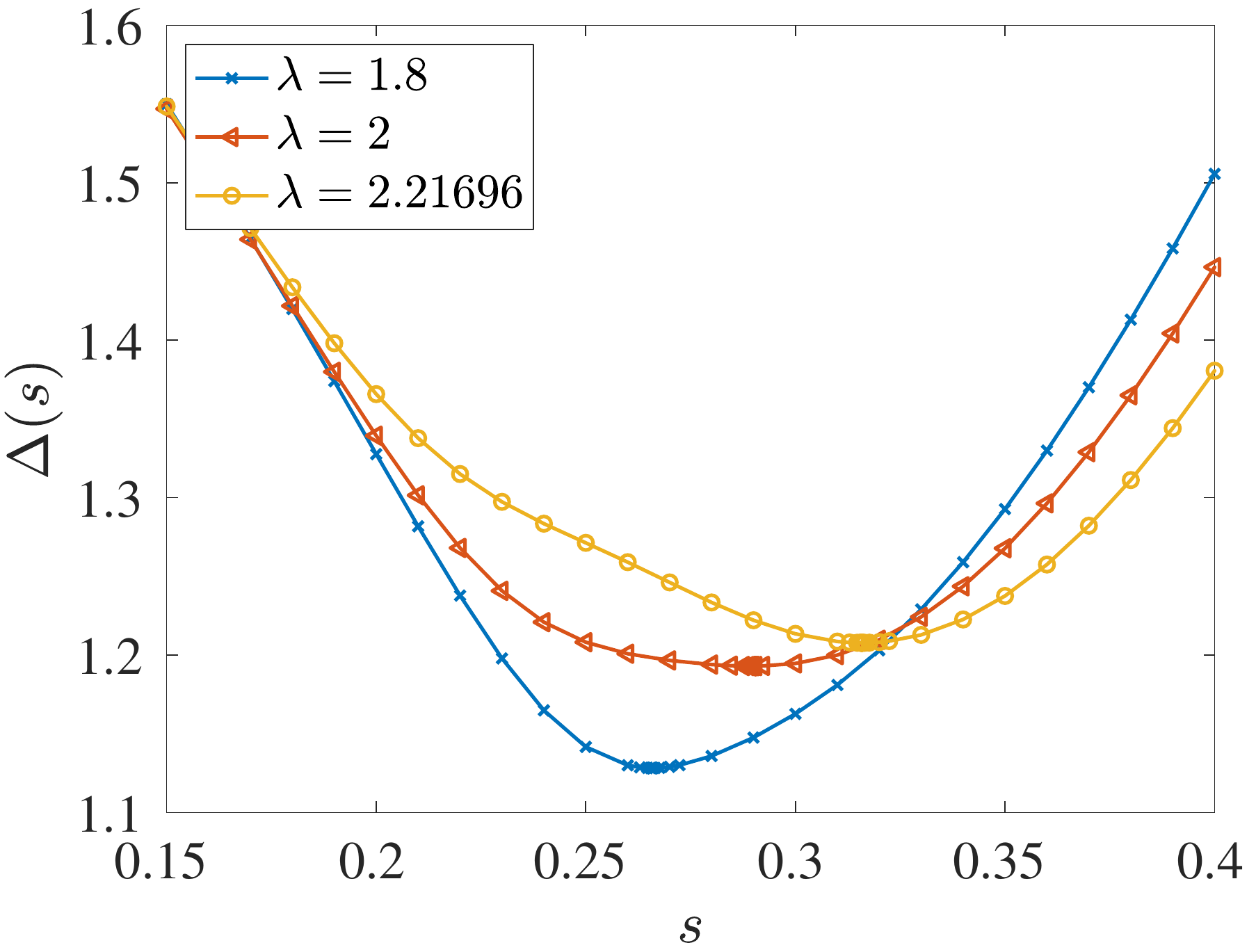}}
   \subfigure[\ $c=0.96$]{\includegraphics[width=0.75\columnwidth]{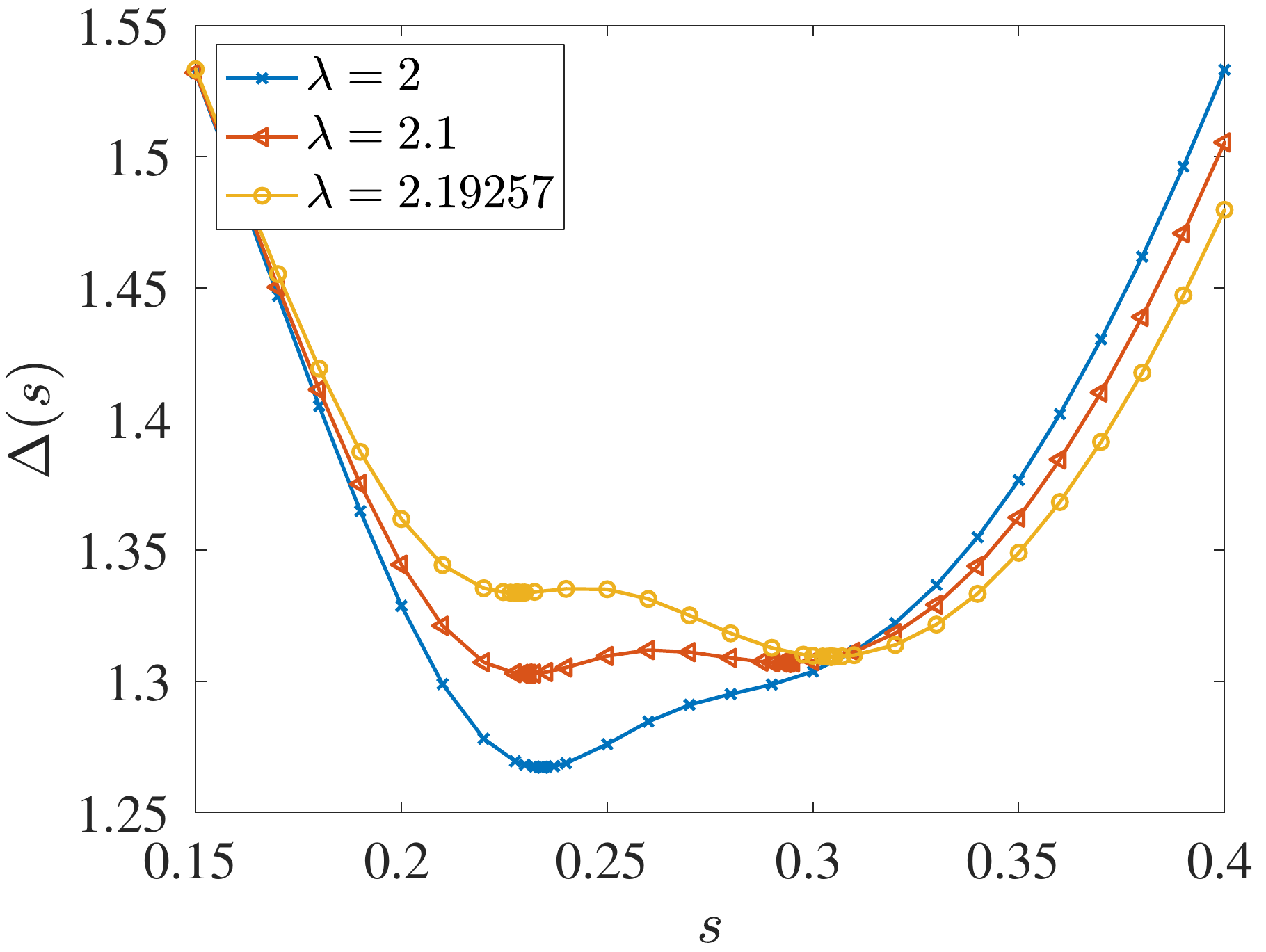}}
   \subfigure[\ $c=0.99$]{\includegraphics[width=0.75\columnwidth]{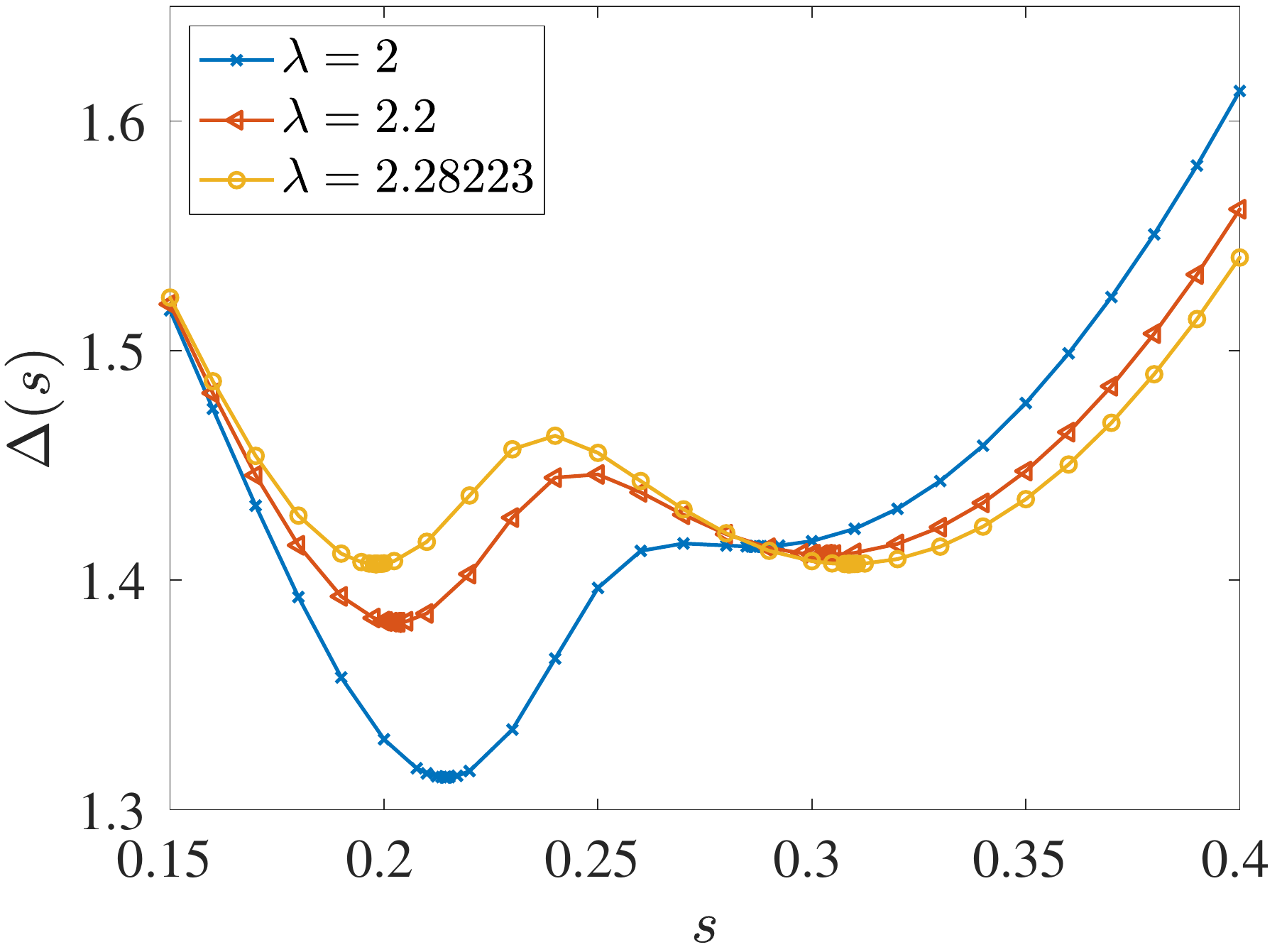}}
   \caption{Behavior of the gap $\Delta(s)$ for the $p=3$ ferromagnetic $p$-spin model in the range of the interpolation where the the minimum gap occurs.}
   \label{fig:FixedpGap2}
\end{figure}

In order to understand the non-monotonic behavior of $\lambda^\ast$ with $c$ observed, it is useful to consider how the gap behaves during the interpolation.  We show in Fig.~\ref{fig:FixedpGap2} the gap behavior at $p=3$ for representative $c$ values.
For $c \lesssim 0.93$, corresponding to the region where $\lambda^\ast$ increases with $c$, the gap exhibits a single minimum along the interpolation, and the minimum grows in value and its location smoothly moves to larger $s$ values as $\lambda$ approaches $\lambda^\ast$. 

For $c \gtrsim 0.93$, corresponding to the region where $\lambda^\ast$ decreases and then increases with $c$, the gap exhibits two minima along the interpolation near $\lambda^\ast$. The first minimum (at the smaller $s$ value) is initially lower in value but grows and passes the second minimum as $\lambda$ approaches $\lambda^\ast$ (the second minimum does not change significantly as $\lambda$ approaches $\lambda^\ast$).  Thus, this region of parameter space is characterized by the global minimum along the interpolation discontinuously jumping in $s$ as $\lambda$ approaches $\lambda^\ast$.  The second minimum (and not the first) then determines the value of $\lambda^\ast$.

\begin{figure}[htbp] 
   \centering
   \subfigure[\ $p=3$]{\includegraphics[width=0.75\columnwidth]{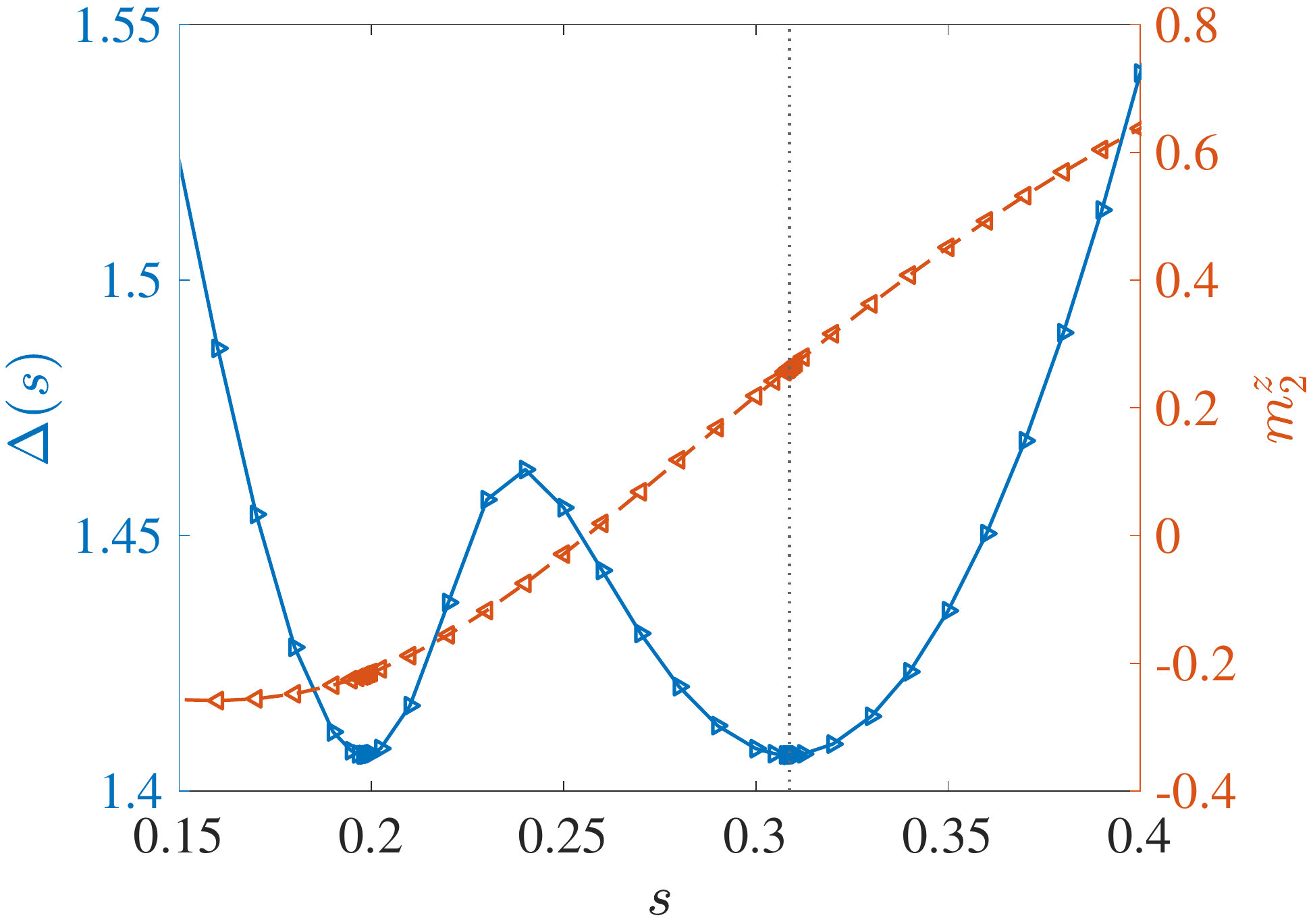}}
    \subfigure[\ $p=5$]{\includegraphics[width=0.75\columnwidth]{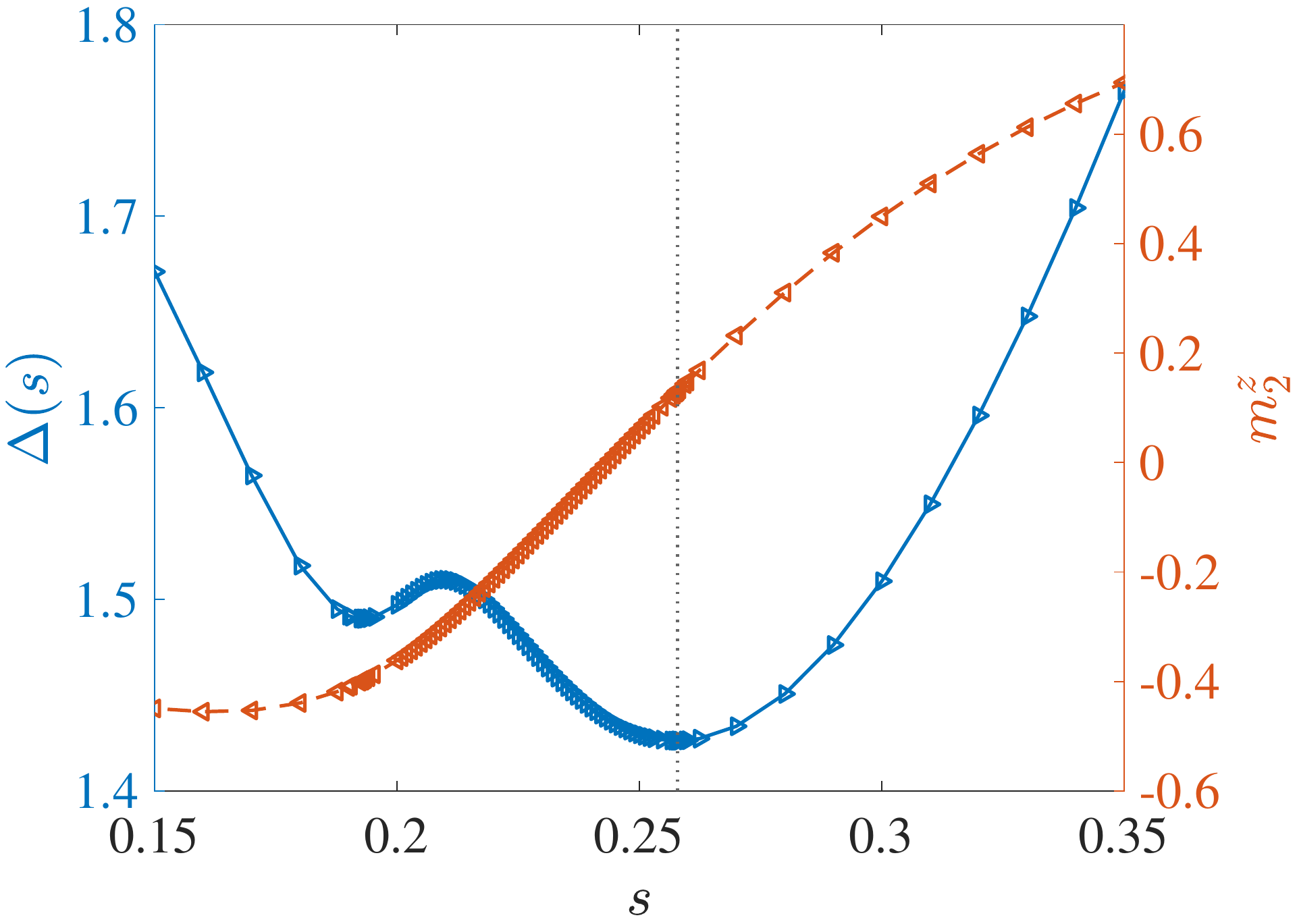}}
     \subfigure[\ $p=7$]{\includegraphics[width=0.75\columnwidth]{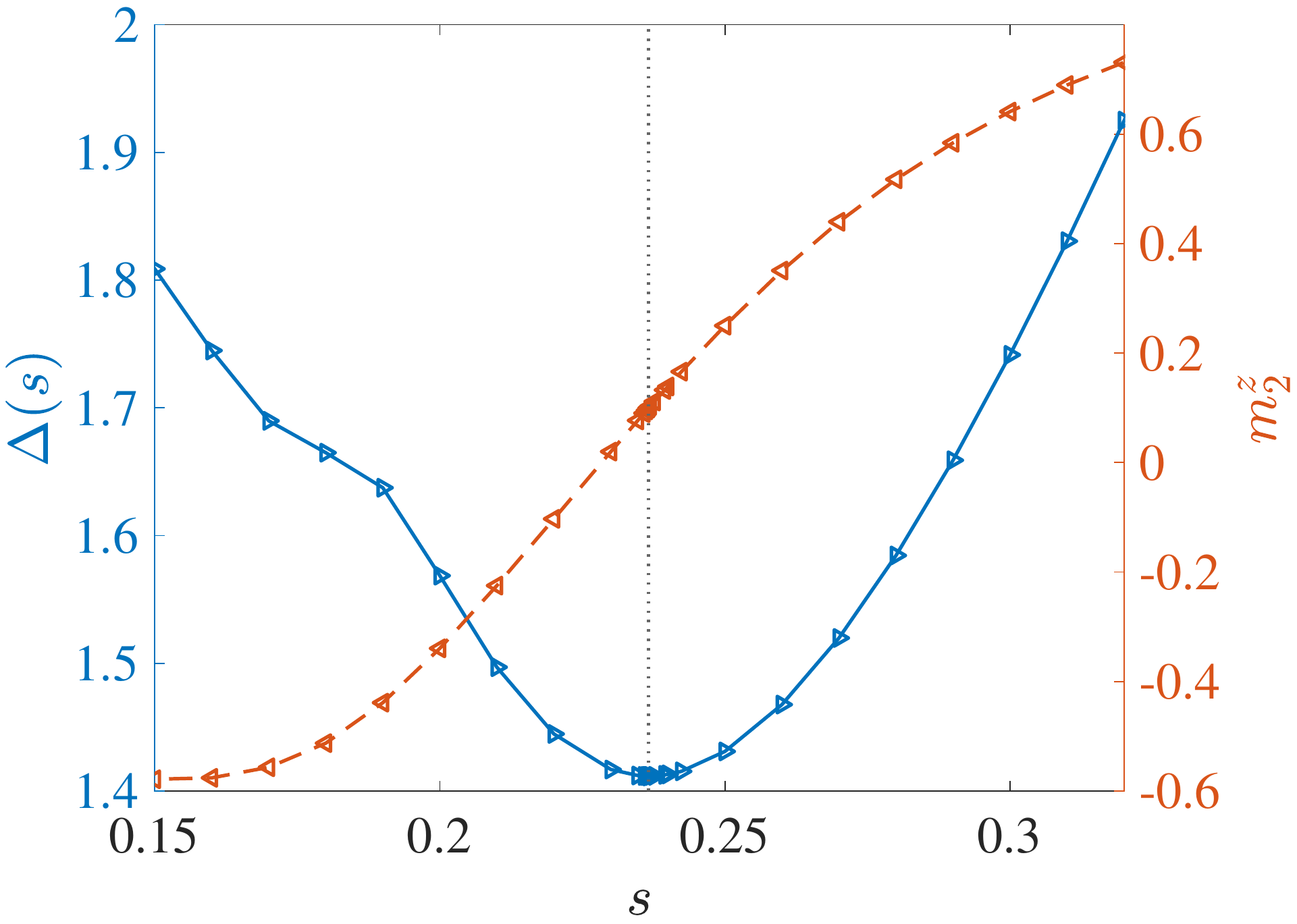}}
   \caption{Behavior of the gap $\Delta(s)$ and $m_2^z$ for $c=0.99$ at the optimal $\lambda = \lambda_\ast$ for the ferromagnetic $p$-spin model. Vertical dotted line corresponds to position of the minimum gap.}
   \label{fig:M2Z}
\end{figure}

The second minimum occurs along the steepest portion of the rise of $m_2^z$ from its negative to positive value (Fig.~\ref{fig:M2Z}), so we can qualitatively associate the second minimum with the flipping of spins associated with the second cluster.  Therefore, at sufficiently high $c$ values, the optimum $\lambda$ is associated with optimizing the hardness associated with flipping the misaligned small cluster. 

\section{Another example of inducing a perturbative crossing} \label{app:LoopGadget3}
We consider a problem Hamiltonian $H_{\mathrm{P}}$ of the form:
\beq \label{eqt:LoopGadget3}
H_{\mathrm{P}}  = \frac{1}{R} \left( \sum_{i=0}^{n-1} h_i \sigma^z_{i} -\sum_{i=0}^{n-1} J_{i,i+1} \sigma_i^z \sigma_{i+1}^z \right) \ ,
\eeq
with $\sigma_n^z \equiv \sigma_0^z$ (periodic boundary conditions). We restrict ourselves to even length chains, but we distinguish between two types of chains.  For $n = 4 k , \ k =  2, 3, \dots$, we choose:
\bes
\begin{align}
J_{i, i+1} =&  \left\{ \begin{array}{ll}
\frac{R}{2} -1 \ , & \text{if } i = \frac{n}{2} , n - 1 \\
R \ , & \text{otherwise}
\end{array} \right. \\
h_i = & \left\{ \begin{array}{ll}
-1 \ , & \text{if } i = \frac{n}{2}+1, \frac{n}{2}+2, \dots, n -1 \\
1 \ , & \text{otherwise}
\end{array} \right. \ ,
\end{align}
\ees
while for $n = 4 k + 2 , \ k = 1, 2, \dots$, we choose:
\bes
\begin{align}
J_{i, i+1} =&  \left\{ \begin{array}{ll}
\frac{R-1}{2}  \ , & \text{if } i = \frac{n}{2}-1 , n-1\\
R \ , & \text{otherwise}
\end{array} \right. \\
h_i = & \left\{ \begin{array}{ll}
-1 \ , & \text{if } i = \frac{n}{2}, \frac{n}{2} + 1, \dots, n-1 \\
1 \ , & \text{otherwise}
\end{array} \right. \ .
\end{align}
\ees
We take $R = n/2$, which for our system sizes is always a positive integer.

The ground state of this Ising system is the state $\ket{\phi} \equiv \ket{0_{n-1} 0_{n-2} \dots 0_{2k+1} 1_{2k} 1_{2k-1} \dots 1_0}$, where the subscript denotes the qubit index, with energy $E_0$.  The doubly degenerate first excited states are the all-zero and all-one bit strings ($\ket{\eta} = \ket{0_{n-1} \dots 0_0}, \ket{\xi} = \ket{1_{n-1} \dots 1_0}$ with energy $E_1$.  For this problem Hamiltonian, the ground state is at least Hamming distance $n/2 -1$ from the nearest first excited state, while the first excited states themselves are Hamming distance $n$ apart.  The Ising energy gap is given by $\Delta_0 = E_1 - E_0 = 2/R = 4/n$

Let us now consider the spectrum of the diagonal catalyst Hamiltonian near $s = 1$. The Hamiltonian at first order in the perturbative parameter $\Gamma = 1-s$ is given by Eq.~\eqref{eqt:HfirstOrder}.  At this order in perturbation theory, the ground state energy is modified to
\beq 
E_0^{(1)}(\Gamma) = E_0 - \Gamma \left( E_0 + \varepsilon_n \right) \ ,
\eeq
 where $\varepsilon_n = 0$ if $n = 4 k + 2$ and $\varepsilon_n = 2$ if $n = 4 k$.
In the first excited state subspace, the first order perturbation is given by
 \beq
V_1 =  \left( H_{\mathrm{D}} + H_{\mathrm{B}} - H_{\mathrm{P}} \right) \rightarrow \left( \begin{array}{cc}
-E_1 + n & 0 \\
0 & -E_1 - n
\end{array} \right)_{\ket{\eta},\ket{\xi}} \ , 
\eeq
where the state $\ket{\xi}$ is lowered in energy by $H_\mathrm{B}$ while the state $\ket{\eta}$ is raised in energy by $H_{\mathrm{B}}$.  Therefore the instantaneous first excited state energy is given by:
\beq
E_1^{(1)}(\Gamma) = E_1 - \Gamma \left( E_1 + n \right) \ .
\eeq
The instantaneous energy gap at first order in the perturbation $\Gamma$ is then given by:
\beq
\Delta_{\mathrm{DC}}^{(1)}(\Gamma) = \Delta_0 - \Gamma (\Delta_0 + n - \varepsilon_n) \ .
\eeq
Because the rate of decline of the instantaneous first excited state is faster than the instantaneous ground state, we predict a perturbative crossing to occur at $\Delta_{\mathrm{DC}}^{(1)}( \Gamma_{\mathrm{DC}}^*)= 0$:
\beq
 \Gamma_{\mathrm{DC}}^* = \Delta_0 /  \left(\Delta_0 + n - \varepsilon_n \right) \ .
 \eeq
This crossing gets closer and closer to $\Gamma = 0$ as $n$ gets larger. Because the ground state and the first excited states are at least $n/2 - 1$ Hamming distance apart, we can expect the avoided level gap at this perturbative crossing to scale as $(\Gamma^\ast_{\mathrm{DC}})^{n/2} \sim  n ^{-n/2}$.  This is in contrast to the standard forward anneal protocol, where the spectrum does \emph{not} exhibit a perturbative crossing at first order in perturbation theory near $\Gamma = 0$. 

\end{document}